\documentclass[12pt]{iopart}
\usepackage[dvips]{epsfig}
\usepackage[usenames]{color}
\usepackage{pstricks}
\usepackage{amssymb}
\usepackage{subfigure}
\usepackage{afterpage}
\usepackage[utf8]{inputenc}
\usepackage{bm}

\begin{document}

\title{Dipolar confinement-induced molecular states in harmonic waveguides}

\author{Gaoren Wang$^1$, Panagiotis Giannakeas$^2$, Peter Schmelcher$^{1,3}$}

\address{$^1$ Zentrum f\"{u}r Optische Quantentechnologien, Universit\"{a}t Hamburg, Luruper Chaussee 149, 22761 Hamburg, Germany}

\address{$^2$ Department of Physics and Astronomy, Purdue University, West Lafayette, Indiana 47907, USA}

\address{$^3$ The Hamburg Centre for Ultrafast Imaging, Universit\"{a}t Hamburg, Luruper Chaussee 149, 22761 Hamburg, Germany}
\vspace{10pt}

\begin{abstract}
The bound states of two identical dipoles in a harmonic waveguide are investigated. 
In the regime of weak dipole-dipole interactions, the local frame transformation (LFT) method is applied to determine the spectrum of dipolar confinement-induced bound states analytically. The accuracy of the LFT approach is discussed by comparing the analytical results with the numerical ones based on a solution of the close-coupling equations. It is found that close to the threshold energy in the waveguide, the LFT method needs to include more partial wave states to obtain accurate bound state energies. As the binding energy increases, the LFT method using a single partial wave state becomes more accurate. We also compare the bound states in waveguides and in free space. For the bosonic case, the $s$-wave dominated bound state looks like a free-space state when its energy is below a certain value. For the fermionic case, the $p$-wave dominated bound state energies in waveguides and in free-space coincide even close to zero energy.
\end{abstract}

%
%
%
%
%

\section{Introduction}
Ultracold gases in tightly confining traps have attracted much attention particularly since one can realize effective one- \cite{prl:94:210401,prl:110:203202} and two- \cite{prl:105:030404} dimensional systems by tuning their geometry. The tight confinement modifies significantly the interparticle collisional properties, and specifically leads to confinement induced resonances (CIRs) which have been predicted theoretically \cite{prl:81:938,prl:84:2551,prl:91:163201} and observed experimentally \cite{prl:94:210401,prl:104:153203}. The capability to tune the two-body interaction via Feshbach resonances and/or CIRs enabled the investigation of strongly correlated many-body physics in low dimensions \cite{science:325:1224}. Tight traps also affect the two-body bound state properties, and confinement induced molecular (CIM) states have been observed for isotropic interparticle interaction \cite{prl:94:210401,prl:110:203202}.\\
\indent Dipolar gases \cite{rpp_72_126401,science:322:231,prl:116:205303,prl:94:160401,prl:112:010404,jpb:49:152001} possess anisotropic dipole-dipole interaction (DDI) \cite{njp:18:113004} which makes it interesting to investigate the influence of a tight trap on such systems. It has been demonstrated that adding a trap in one direction can suppress the reactive scattering collisions in a polar molecular gas \cite{np:7:502}. In addition, one can control the reactive collisions by tuning the orientation of the dipoles with respect to the confined direction \cite{pra:92:042706}. The Influence of a two-dimensional trap on the dipolar reactive collision has been studied in Refs. \cite{njp:17:013020,njp:17:035007}. In the non-reactive case, the modification of the DDI by traps has been analyzed theoretically in \cite{prl:99:140406,pra:89:023604} thereby demonstrating that dipolar confinement induced resonances (DCIR) occur in harmonic waveguides \cite{prl:111:183201}. For the case that the confining potential is anharmonic, inelastic DCIRs have been predicted \cite{njp:17:065002}. The relative orientation between dipoles and the waveguide axis can also be used to tune the two-body interaction in a quasi-one dimensional geometry \cite{pra:90:042710}. While the scattering properties of two dipoles in traps have been theoretically investigated, the dipolar CIM (DCIM) state properties of such system have been addressed much less. \\
\indent In this work, we consider two identical dipoles, which can be either bosonic or fermionic, in a harmonic waveguide. The dipoles are aligned along the longitudinal direction ($z$ axis) of the waveguide. The local frame transformation (LFT) approach \cite{prl:111:183201,pra:24:619,prl:49:128,pra:36:4236,pra:92:022711,pra:91:043424,pra:94:013419,prl:92:133202,pra:86:042703,pra:89:052716,pra:92:022706,pra:88:012715,jpb:49:165302} is applied to calculate the DCIM states when the DDI is weak. A dipolar bound state equation is derived which allows one to determine the energies of the DCIM states analytically with the free-space scattering information as input. Moreover, the dipolar bound state equation within the Born approximation shows explicitly the influence of the DDI in determining the bound state energies. By comparing the LFT results with corresponding numerical calculations, it is found that, below threshold, the LFT approach with the single partial wave approximation is accurate even in the presence of DDI. Close to threshold, one needs to include higher partial wave states in the LFT approach to get accurate bound state energies. The dependence of the DCIM states on the DDI strength is explored. Based on numerical calculations, both the weak and strong DDI regimes are investigated. We find that qualitatively the dependence of the DCIM states on the DDI is similar in these two regimes. The DCIM state becomes increasingly bound as the DDI increases. The $l>0$-wave dominant bound states are more sensitive to the variation of the DDI compared to the $s$-wave dominant states. New DCIM states can emerge by increasing the DDI. \\
\indent The paper is organized as follows. Section II introduces our computational methods. The set of close coupling equations is provided in a partial wave basis, and the dipolar bound state equations based on the LFT approach are presented. In Sec. III, the properties of the dipolar confinement induced molecular states are discussed. Both the bosonic and fermionic cases are analyzed. Sec. IV contains our conclusions.
\section{Computational method}
In harmonic waveguides, the center of mass motion and relative motion are separable. The Hamiltonian of the relative motion, which contains a short-range isotropic interaction in conjunction with DDI, is expressed as
\begin{eqnarray}
H=T+V_{t}(\bm{r})+V_{\rm{sr}}(\bm{r})+V_{d}(\bm{r}),
\label{eq_Hamiltonian}
\end{eqnarray}
where $T$ is the kinetic energy. $V_{t}(\bm{r})$ is the transverse trapping potential, and is assumed to be an isotropic two-dimensional harmonic potential  $V_{t}(\bm{r})=\frac{1}{2}\mu\omega_{\perp}^2\rho^2$, where $\mu$ is the reduced mass, $\omega_{\perp}$ is the trapping frequency, and $\rho$ is the magnitude of the transverse component of the interparticle vector $\bm{r}$. $V_{\rm{sr}}(\bm{r})$ is the short-range isotropic potential, and depends on the species under consideration. $V_{\rm{sr}}(\bm{r})$ is modeled by a Lennard-Jones (LJ) potential $V_{\rm{sr}}(\bm{r})=C_{10}/r^{10}-C_{6}/r^6$ which possesses a van der Waals potential tail (note that $r=|\bm{r}|$).
$V_{d}(\bm{r})$ is the DDI, and has the usual form $V_{d}(\bm{r})=\frac{d^2}{r^3}(1-3\cos^2\theta)$ where $d$ is the dipole moment, and $\theta$ is the angle between the $z$ axis and the interparticle vector $\bm{r}$. We remark that the singularity of the DDI at the origin is remedied by the corresponding behavior of the short range potential at the origin.  The threshold energy of two dipoles without the confinement is chosen to be the zero energy point. The energy of the scattering threshold in waveguides is given by $E_{\rm th}=\hbar\omega_{\perp}$.\\
\indent The three potential terms in Eq.~(\ref{eq_Hamiltonian}) determine three length scales in the system. The length scale associated with the transverse confinement is the harmonic oscillator length  $a_{\perp}=\sqrt{\hbar/\mu\omega_{\perp}}$ whereas the length scale of the short-range interaction term is the van der Waals length given by the relation $\beta_6=(2{\mu}C_6/\hbar^2)^{1/4}$. Finally the DDI is characterized by the dipole length $l_{d}={\mu}d^2/\hbar^2$.
Below we will introduce two methods to determine the bound state belonging to the Hamiltonian (\ref{eq_Hamiltonian}).
One approach is the close-coupling method which solves the problem numerically.
In the weak DDI regime, the dipole length and the van der Waals length are far smaller than the harmonic oscillator length. The local frame transformation method is in this case applied to derive the DCIM states (semi)analytically.
\subsection{Close-coupling method}
We expand the two-body wavefunction $\psi$ in the partial wave basis
\begin{equation}
\psi(r,\theta,\phi)=\frac{1}{r}\sum_{lm}f_{lm}(r)Y_{lm}(\theta,\phi),
\end{equation}
where $f_{lm}(r)$ are the radial wavefunctions, $Y_{lm}$ are the spherical harmonics, and $l$ and $m$ are the partial wave quantum number and the magnetic quantum number, respectively. Since the system under investigation is cylindrically symmetric, the magnetic quantum number is conserved. In the following $m$ is set to zero, and is consequently omitted. The kinetic term $T$ and the short-range potential $V_{\rm sr}({\bm r})$ are diagonal in the partial wave basis. The dipole potential $V_d({\bm r})$ and the transverse trapping potential $V_{t}({\bm r})$ couple different partial wave states. The matrix elements of $V_d({\bm r})$ and $V_{t}({\bm r})$ in the partial wave basis are given respectively by \cite{njp:11:055039}
\begin{equation}
V_d^{ll^\prime}({\bm r})=<l|V_d|l^\prime>=-\frac{2d^2}{r^3}\sqrt{(2l+1)(2l^{\prime}+1)}
\left(
\begin{array}{c c c}
l & 2 & l^\prime 			\\
0 & 0 & 0					\\
\end{array}
\right)^2,
\end{equation}
\begin{equation}
V_{\rm trap}^{ll^\prime}({\bm r})=<l|V_{\rm trap}|l^\prime>=\frac{1}{3}\mu\omega_{\perp}r^2\delta_{ll^\prime}-\frac{1}{3}\mu\omega_{\perp}r^2\sqrt{(2l+1)(2l^{\prime}+1)}
\left(
\begin{array}{c c c}
l & 2 & l^\prime 			\\
0 & 0 & 0					\\
\end{array}
\right)^2, 
\end{equation}
where $\delta$ is the Kronecker delta function, and the large curved brackets are 3-j symbols.
In the partial wave basis, the Schr\"{o}dinger equation is a set of close-coupling equations satisfied by the radial wavefunction $f_{l}(r)$
\begin{eqnarray}
&\sum_{l}\left[
-\frac{\hbar^2}{2\mu}\frac{d^2}{dr^2}+V_{c}({\bm r})+V_{\rm sr}({\bm r})
\right]f_{l}(r) \nonumber \\
+&\sum_{ll^\prime}\left(V_d^{ll^\prime}({\bm r})+V_{\rm trap}^{ll^\prime}({\bm  r})\right)f_{l^\prime}(r)
=E\sum_{l}f_{l}(r), 
\label{eq_close-coupled-equation}
\end{eqnarray}
where $V_{c}({\bm r})=\frac{l(l+1)}{2{\mu}r^2}$ is the centrifugal term, and $E$ is the total energy. With the boundary condition that the bound state wavefunction vanishes at $r\rightarrow{0}$ and $r\rightarrow{\infty}$, the set of close-coupling equations given in Eq.~(\ref{eq_close-coupled-equation}) is solved numerically based on the log-derivative algorithm \cite{jcp:85:6425}.
To obtain the bound state energy $E_b$ and the wavefunction $\psi$ the approach of Ref.~\cite{cpc:84:1} is employed.\\
\indent In the numerical calculation, the dimensionless version of Eq.~(\ref{eq_close-coupled-equation}) is used, in which the length and energy are scaled by $a_{\perp}$ and $\hbar\omega_{\perp}$ respectively. The coefficient $C_6$ in the LJ potential is fixed, such that $\beta_6/a_{\perp}$ is 0.018, which is an experimentally achievable value. For example, the van der Waals coefficient of two ground state $^{166}$Er atoms is 1723~au \cite{nature:507:7493}, and the corresponding van der Waals length $\beta_6$ is 151~au. In Ref.~\cite{science:345:1484}, a transverse trapping potential with frequency $\omega_{\perp}\sim{600}$~Hz is realized experimentally for Er atoms. The corresponding harmonic oscillator length $a_{\perp}$ is 8518~au, and the ratio $\beta_6/a_\perp$ amounts to 0.018. For other systems, the van der Waals lengths would be different. Nevertheless one can tune the transverse trapping frequency $\omega_{\perp}$ to achieve the desired value for $\beta_6/a_{\perp}$. The coefficient $C_{10}$ in the LJ potential is varied such that the scattering length of the LJ potential can be changed significantly, and moreover, the number of bound states supported by the LJ potential can be tuned.
\subsubsection{Partial wave probability density of the DCIM states}
The partial wave probability densities (PD) of the DCIM state $P_{b}^l=|f_l|^2$ are calculated via close-coupled method \cite{jcp:85:6425,cpc:84:1}. By examining the partial wave PDs, regions dominated by different terms in the Hamiltonian (\ref{eq_Hamiltonian}) are identified.

\begin{figure}
	\centering
	\begin{minipage}[b]{0.5\textwidth}
		\includegraphics[width=\textwidth]{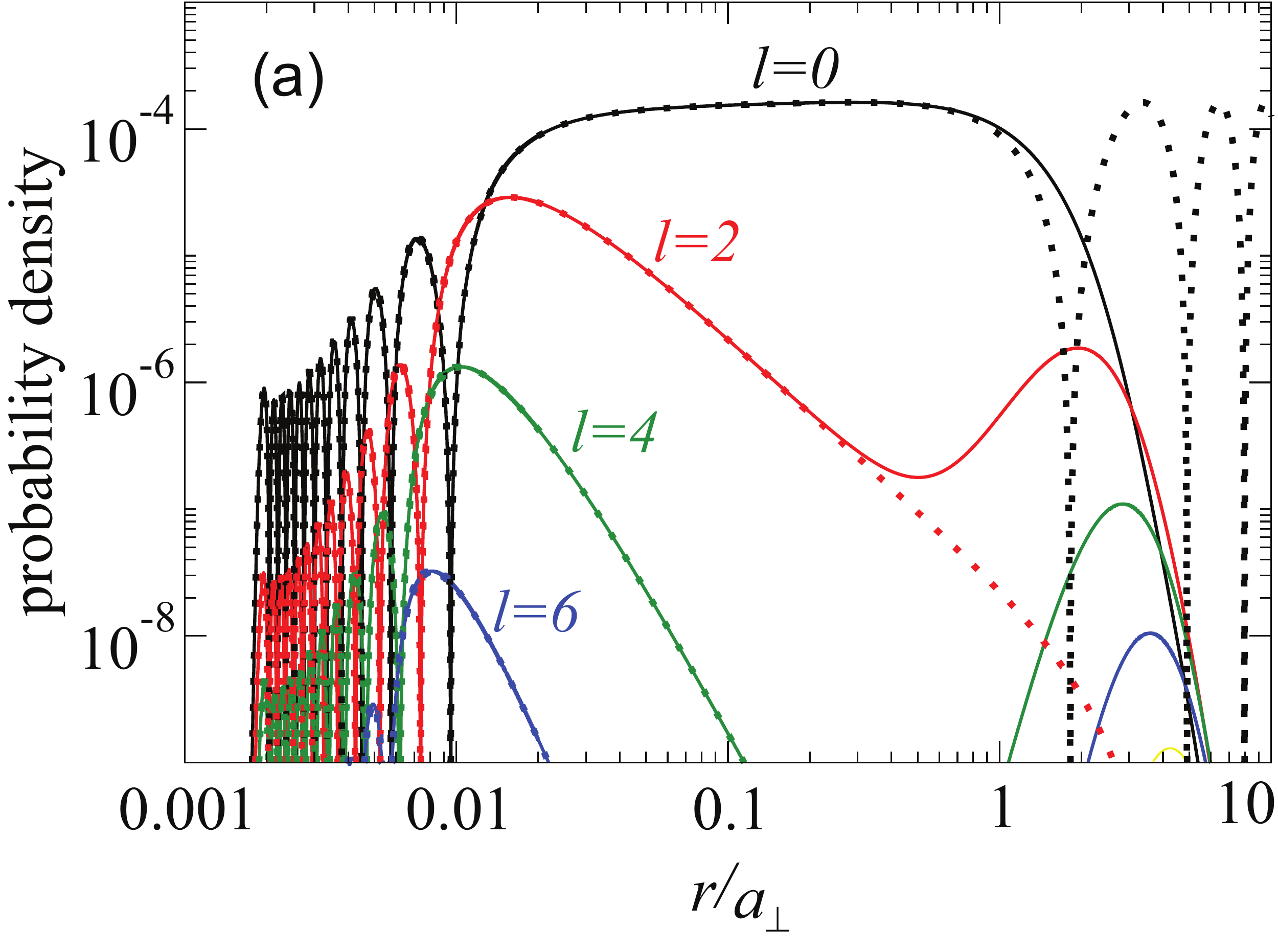} \\
	\end{minipage}
	\begin{minipage}[b]{0.5\textwidth}
		\includegraphics[width=\textwidth]{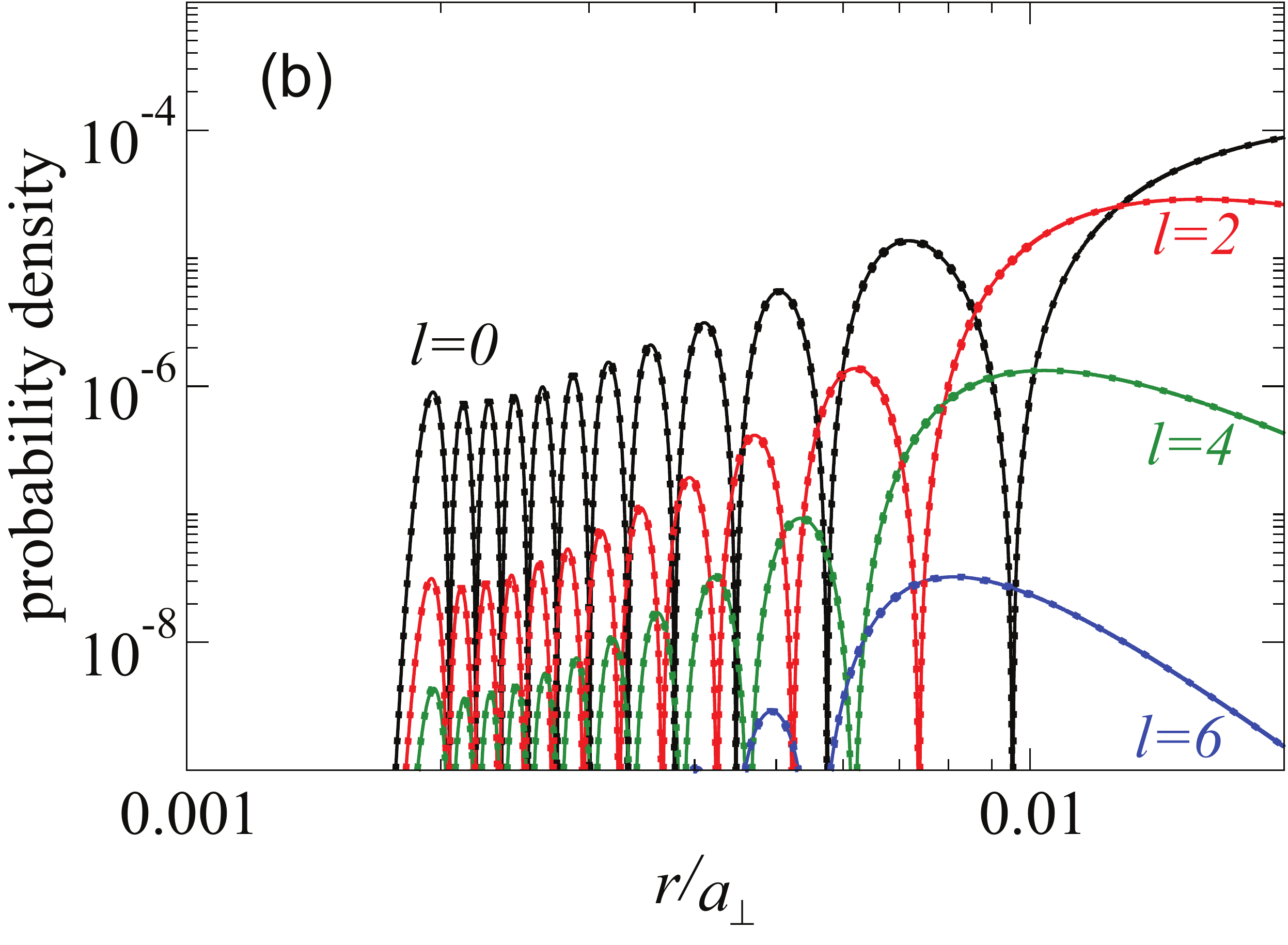}	\\
	\end{minipage}
	\caption{(Upper panel) Bound state partial wave probability density $P_{b}^{l}$ of two identical bosonic dipoles in a waveguide is shown (solid line). The free-space scattering state probability density $P_{s}^{l}$ is also provided (dashed line). $C_{10}^{s}=6.7\times{10^{-19}}$ and the LJ potential supports ten bound states. The ratio $l_d/a_{\perp}$ is 0.026. The bound state energy and the scattering energy are the same $E/\hbar\omega_{\perp}=0.5$. $P_{s}^{l}$ is renormalized at $r_0/{a_{\perp}}=0.01$, so that $P_{s}^{l=0}(r_0/a_{\perp})=P_{b}^{l=0}(r_0/a_{\perp})$. (Lower panel) A zoom-in picture at short distances.}
	\label{fig_wavefunction}
\end{figure}

The partial wave PDs for a bosonic DCIM state are shown in the upper panel of Fig.~\ref{fig_wavefunction} (solid line). The energy of the DCIM state is $E_b/\hbar\omega_{\perp}=0.5$. The scaled $C_{10}^{s}=C_{10}/(\hbar\omega_{\perp}a_{\perp}^{10})$, which is dimensionless, is set to be 6.7$\times{10^{-19}}$. The ratio $l_d/a_{\perp}$ is 0.026. In the range $r/a_{\perp}{\ll}1$, the trapping potential is far smaller than the short-range potential and DDI, and can be neglected. As a demonstration, we performed the free space scattering calculation by dropping the term $V_{t}({\bm r})$ in Eq.~(\ref{eq_Hamiltonian}). The partial wave PD $P_{s}^l$ of the scattering state at energy $E_b$ is also shown in the upper panel of Fig.~\ref{fig_wavefunction} (dotted line). The lower panel of Fig.~\ref{fig_wavefunction} is a zoom-in plot of the upper panel in the short-range region $r/a_{\perp}{\ll}1$, and clearly shows that $P_{b}^l$ has the same nodal structure as $P_{s}^l$ in this region. This indicates that the trapping potential is negligible for $r/a_{\perp}{\ll}1$, and the two dipoles interact essentially like in free space.

\begin{figure}[!h]\centering
	\resizebox{0.5\textwidth}{!}{
		\includegraphics{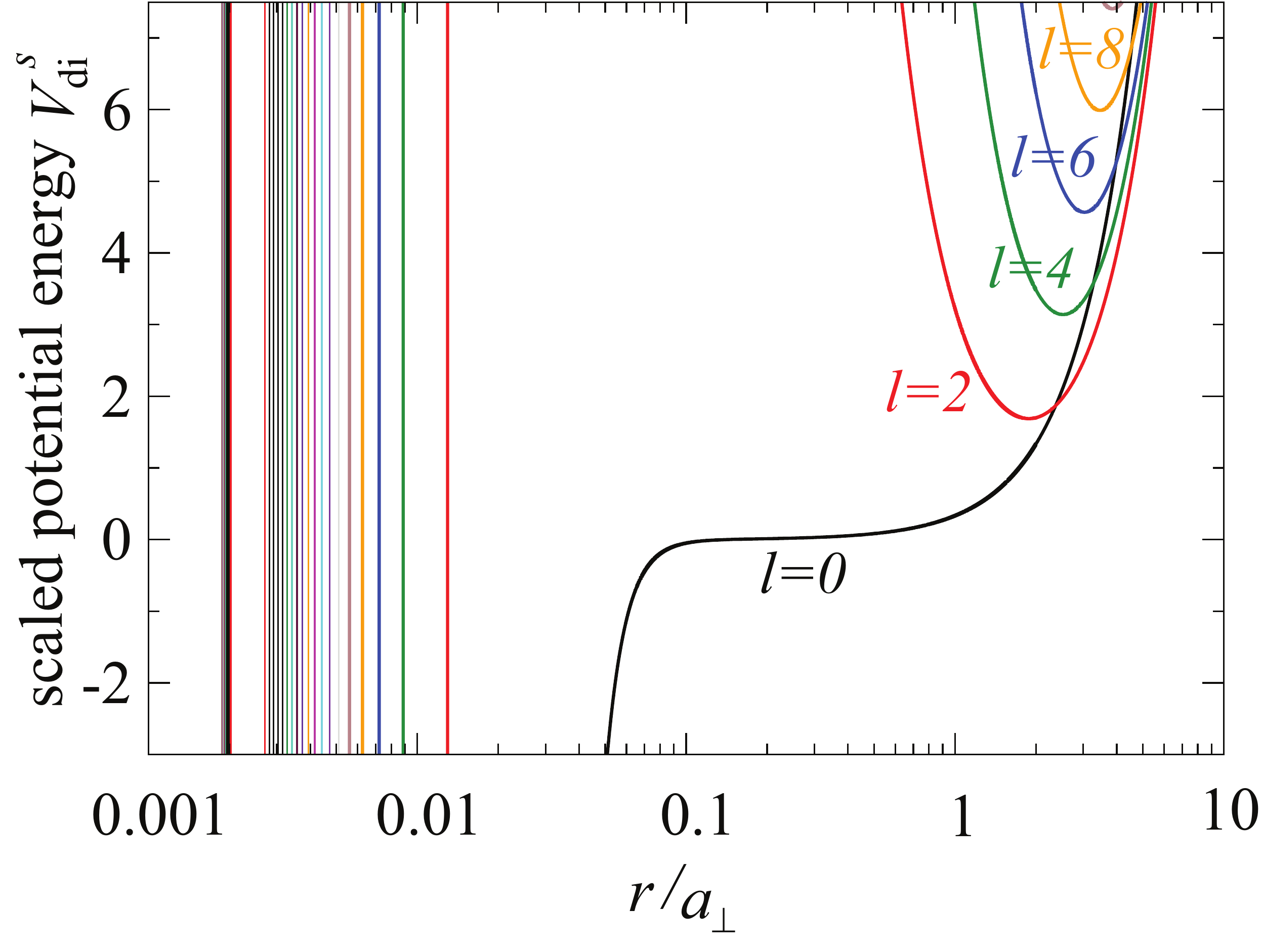}
	} \caption{ The scaled diagonal potential energy $V_{\rm di}^s({\bm r})=V_{\rm di}({\bm r})/\hbar\omega_{\perp}$ in the partial wave basis with $V_{\rm di}({\bm r})=V_{\rm sr}({\bm r})+V_d({\bm r})+V_{t}({\bm r})+V_{c}({\bm r})$. The partial wave quantum number $l$ is associated to the corresponding channel potential.
}\label{fig:diabaticPotential}
\end{figure}

The partial wave PDs shown in Fig.~\ref{fig_wavefunction} offer more information.
The DCIM state (solid line in Fig.~\ref{fig_wavefunction}) involves many partial wave components, and for the depicted state, the dominant one is the $s$-wave component. The oscillatory behavior of $P_{b}^l$ in the short-range region $r/a_{\perp}{\ll}1$, shown in the lower panel of Fig.~\ref{fig_wavefunction}, is due to the presence of the potential well of the interaction potential $V_{\rm int}({\bm r})=V_{\rm sr}({\bm r})+V_{d}({\bm r})$. It is noted that there are maxima for $P_{b}^l$'s with $l>0$ at large distances $r/a_{\perp}>1$.
By examining the potential curves in the partial wave basis, we encounter potential wells at distances $r/a_{\perp}>1$ for $l>0$ partial wave channels as shown in Fig.~\ref{fig:diabaticPotential}. Considering weak DDI, the potential well results from the competition between the decreasing centrifugal potential $V_{\rm cent}$ and the increasing trapping potential $V_{\rm trap}({\bm r})$ as the interparticle distance increases. The positions of the maxima for the high partial wave PDs coincide with the positions of the minima of the outer potential wells at $r/a_{\perp}>1$.

\begin{figure}[!h]\centering
	\resizebox{0.5\textwidth}{!}{
		\includegraphics{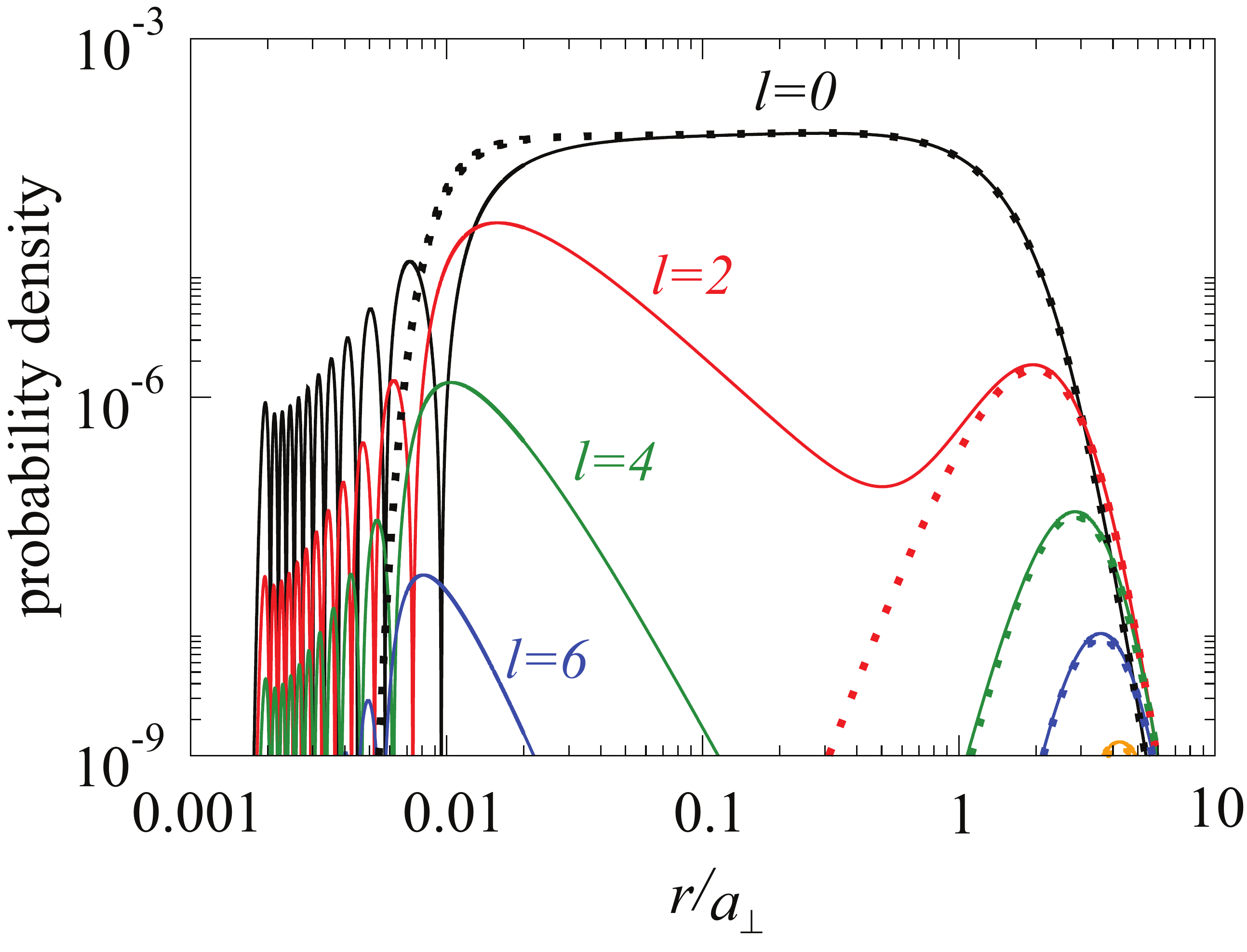}
	} \caption{ The bound state probability density $P_{b}^{l}$ shown in Fig.~\ref{fig_wavefunction} is given via the solid line. The bound state probability density $P_{b}^{\prime{l}}$ (dotted line) is calculated for $C_{10}^s=2.18{\times}10^{-16}$ and $l_d/a_{\perp}=0$. The corresponding LJ potential supports one bound state in free space. The two bound states possess the same energy.
}\label{fig:wv_boundVSbound}
\end{figure}

In the region $r/a_\perp>1$, the short-range potential $V_{\rm sr}({\bm r})$ and the DDI potential $V_d({\bm r})$ decay to zero, and the trapping potential $V_{\rm trap}({\bm r})$ becomes dominant. In this region the system is mainly governed by $V_{t}({\bm r})$, and is nearly independent of $V_{\rm sr}({\bm r})$ and $V_{d}({\bm r})$. In order to confirm this, we compare in Fig.~\ref{fig:wv_boundVSbound} the partial wave PDs of two DCIM states calculated with different short-range potential $V_{\rm sr}({\bm r})$ and DDI potential $V_d({\bm r})$.
The bound state PD $P_{b}^{l}$ shown in Fig.~\ref{fig_wavefunction} is also shown in Fig.~\ref{fig:wv_boundVSbound} (solid line). As stated before, in the calculation of $P_{b}^l$ the LJ potential supports ten bound states, and the dipole moment $l_d/a_{\perp}$ is 0.026. Another bound state PD $P_{b}^{\prime{l}}$, the energy of which is also $E_b$, is additionally shown (dotted line). The LJ potential supports here one bound state and $l_d/a_{\perp}$ is zero in the calculation of $P_{b}^{\prime{l}}$. Due to the different $V_{\rm sr}({\bm r})$ and $V_{d}({\bm r})$ used in the calculations, the short-range parts of the two bound state PDs are significantly different. Nevertheless, in the region $r/a_{\perp}>1$, the two bound state PDs are nearly the same, and are, to a large extend, determined by the trapping potential.  \\
\indent Based on the above observation of length scale separation in the system, we introduce the local frame transformation approach below, which can connect the two-body properties in waveguides with the scattering properties in free space analytically. \\

\subsection{Local frame transformation approach}
The concept of the LFT approach was introduced in Ref. \cite{pra:24:619,prl:49:128} to calculate the Stark effect of nonhydrogenic Rydberg spectra. Subsequently the method was generalized to study the photodetachment of negative ions in magnetic fields \cite{pra:36:4236,pra:92:022711} and the photoionization of atoms \cite{pra:91:043424,pra:94:013419}. The application of the LFT approach to ultracold collisions in quasi-low dimensional geometry was pioneered in Ref. \cite{prl:92:133202}. Until now, it has been applied to understand different aspects of ultracold collisions in harmonic waveguides, such as higher partial wave confinement-induced resonances(CIR) \cite{pra:86:042703}, energy dependence of the CIRs \cite{pra:89:052716}, multi-open-channel collisions \cite{pra:92:022706} and the dipolar CIRs \cite{prl:111:183201}. Ultracold collisions in other confining geometries have been discussed in \cite{pra:88:012715}. Recently the LFT approach has been adopted to treat the two-body scattering analytically in the presence of spin-orbit coupling \cite{arxiv:1701}. Here we provide a further application of the LFT approach, and show that one can obtain a comprehensive analysis of the DCIM states. In the following, the key idea of the LFT approach is briefly introduced (see Ref.\cite{prl:92:133202,pra:86:042703,pra:89:052716} for more details) and then the bound state equation for two dipoles in waveguides is presented. Finally, within the Born approximation, the dipolar bound state equation is simplified to show the dependence on the DDI explicitly.
\subsubsection{LFT method for bound states: brief review}
As shown in Fig.~\ref{fig_wavefunction} and \ref{fig:wv_boundVSbound}, the length scales of the short-range potential $V_{\rm sr}({\bm r})$, DDI potential $V_d({\bm r})$ and the trapping potential $V_{t}({\bm r})$ are separated in the weak DDI regime. In the region $r/a_{\perp}\ll{1}$, $V_{t}({\bm r})$ is negligibly small compared to $V_{\rm sr}({\bm r})$ and $V_d({\bm r})$, and the two dipoles effectively interact as if there is no confinement. The partial wave basis is employed to construct the wavefunction in this region. The DDI couples different partial wave states, and the effect of the two-body interaction is encapsulated in the free space K matrix, denoted as $\bf{K}^{3D}$. In the region $r/a_{\perp}{\gg}1$, the interparticle interaction $V_{\rm sr}({\bm r})+V_d({\bm r})$ vanishes, and the system can be treated as two non-interacting particles in a waveguide. 
The wavefunction in this region can be written as 
\begin{equation}
\Psi(\bm r)=\bm{F}-\bm{G}\bm{K}^{1D}, 
\label{eq_wv_asymptotic}
\end{equation}
where $\bm{F}={\rm diag}\{F_1,F_2,\cdots\}$ and $\bm{G}={\rm diag}\{G_1,G_2,\cdots\}$ are diagonal matrices. $F_n$ and $G_n$ are the regular and irregular solutions of the Hamiltonian (\ref{eq_Hamiltonian}) without the short-range potential $V_{\rm sr}({\bm r})$ and the DDI potential $V_d({\bm r})$. The explicit expressions for $F_n$ and $G_n$ have been given in \cite{pra:86:042703} which are 
the product of the eigenfunctions of the two dimensional harmonic oscillator in the transverse plane and the standing wave with proper symmetry in the longitudinal direction.
$\bf{K}^{1D}$ is the K matrix in such quasi one-dimensional geometry. In the intermediate region $\beta_6/a_{\perp}, l_d/a_{\perp}<r<1$, both $V_{\rm sr}({\bm r})+V_d({\bm r})$ and $V_{\rm trap}({\bm r})$ are small compared to the kinetic energy. Then in this region, both the partial wave basis and the asymptotic basis, ie. $F_n$ and $G_n$, can be used to describe the wavefunction. A local transformation matrix $\bf{U}$ can be defined which connects the two basis sets. The element of the transformation matrix $\bf{U}$ reads \cite{pra:89:052716,jpb:49:165302}
\begin{equation}
U^T_{l, n}=\frac{\sqrt{2}(-1)^{d_0}}{a_{\perp}}\sqrt{\frac{2l+1}{kq_n}}P_l\left(\frac{q_n}{k}\right),
\label{eq_LocalTransformationMatrixElement}
\end{equation}
where $n$ is the transverse harmonic oscillator quantum number. $d_0$ is $l/2$ for even $l$ and $(l+1)/2$ for odd $l$. $P_l(x)$ is Legendre polynomial, $q_n$ is the channel momentum along the waveguide axis determined by $\frac{(\hbar{q_n})^2}{2\mu}=E-\hbar\omega_{\perp}(2n+1)$.

The local frame transformation $\bf{U}$ can be used to express $\bf{K}^{1D}$ in terms of $\bf{K}^{3D}$ according to the relation \cite{pra:89:052716}
\begin{equation}
\bf{K}^{1D}=\bf{U}\bf{K}^{3D}\bf{U}^T.
\label{eq_K1D}
\end{equation}
From the $\bf{K}^{1D}$ matrix, one can deduce both bound state and scattering information in waveguides \cite{pra:89:052716}. We are interested in the bound state spectrum here. By imposing asymptotically an exponentially decaying boundary condition in the wavefunction (\ref{eq_wv_asymptotic}), one obtains the following relation for $\bf{K}^{1D}$ \cite{pra:86:042703,pra:89:052716}
\begin{equation}
\det(\bf{I}-i\bf{K}^{1D})=0,
\label{eq_boundlevelinwaveguide}
\end{equation}
where the roots of Eq.~(\ref{eq_boundlevelinwaveguide}) provides us with the energies of the confinement-induced bound state $E_b$.
\subsubsection{Dipolar bound state equation}
\indent Next we examine the explicit expression for the $\bf{K}^{3D}$ matrix in the presence of DDI, and derive the dipolar bound state equation in terms of the  $\bf{K}^{3D}$ matrix. For systems consisting of atoms governed by the van der Waals interaction, the single partial wave approximation works quite well in the ultracold regime \cite{pra:89:052716}. In the presence of DDI, different partial wave states are coupled \cite{pra:89:022702}. To be specific, the $l$ partial wave is coupled to the $l^{\prime}$ partial wave with $l^{\prime}=l,l\pm2$. $l=l^{\prime}=0$ is an exceptional case which is not coupled by the DDI. In the determination of DCIR \cite{prl:111:183201}, the LFT approach with three partial wave states can accurately reproduce the numerical results in the weak DDI regime. Therefore, we include up to three lowest partial wave states in the derivation of the bound state equation either for bosons or fermions.
For identical particles, the free space $K^{3D}$ matrix including three partial wave states can be expressed as \cite{prl:111:183201}
\begin{equation}
K^{3D}=
\left(
\begin{array}{c c c}
K_{l_1,l_1} & K_{l_1,l_2} & 0 			\\
K_{l_2,l_1} & K_{l_2,l_2} & K_{l_2,l_3}	\\
0			& K_{l_3,l_2} & K_{l_3,l_3} \\
\end{array}
\right),
\label{eq_K3D}
\end{equation}
where $l_1$, $l_2$, and $l_3$ label the quantum numbers of the lowest three partial wave states in ascending order. For identical bosons, the lowest three partial wave states are the $s$, $d$ and $g$ wave states. For identical fermions, these are the $p$, $f$ and $h$ wave states. In the expression of $K^{3D}$ (see Eq.~(\ref{eq_K3D})),
direct couplings between different partial waves due to the DDI are included, such as $K_{l_1,l_2}$, $K_{l_2,l_1}$ and $K_{l_2,l_3}$, $K_{l_3,l_2}$. Since the weak DDI is considered here, the indirect couplings between partial wave states which are mediated by another state are set to zero, such as $K_{l_3,l_1}$ and $K_{l_1,l_3}$.

From the $K^{3D}$ matrix, the generalized scattering length $a_{l,l^{\prime}}$ is introduced as \cite{pra:75:052705}
\begin{equation}
a_{l,l^{\prime}}=-K_{l,l^{\prime}}/k,
\end{equation}
where $k=\sqrt{2\mu{E}/\hbar^2}$ is the collisional momentum. By substituting Eqs.~(\ref{eq_K1D}) and (\ref{eq_K3D}) into Eq.~(\ref{eq_boundlevelinwaveguide}), the dipolar bound state equation including three partial wave states can be written as
\begin{equation}
a_{l_1,l_1}=\frac{i}{k}\frac{\Delta_{N}}{\Delta_{D}},
\label{eq_BoundStateEquation_threeL}
\end{equation}
where
\begin{eqnarray}
\Delta_{N}=&-1+2 i M_ {l_3,l_2} K _{l_3,l_2}+M_ {l_3,l_2}^2
K _ {l_3,l_2}^2+i M_ {l_3,l_3} K _{l_3,l_3}-M_ {l_3,l_2}^2 K
_ {l_2,l_2} K _{l_3,l_3} \nonumber \\
&-2 M_{l_3,l_2} M_ {l_3,l_1}
K _ {l_2,l_1} K _{l_3,l_3}-i M_ {l_3,l_2}^2 M_ {l_1,l_1} K _ {l_2,l_1}^2
K _{l_3,l_3}+M_ {l_2,l_1}^2 K _ {l_2,l_1}^2 \left(1-i
M_ {l_3,l_3} K _{l_3,l_3}\right) \nonumber \\
&+M_ {l_2,l_2} (-2 M_{l_3,l_1}
K _ {l_2,l_1} K _{l_3,l_2}-M_ {l_3,l_3} K _{l_3,l_2}^2-i M_ {l_3,l_1}^2 K _ {l_2,l_1}^2 K _{l_3,l_3} \nonumber \\
&+M_ {l_1,l_1}
K _ {l_2,l_1}^2 \left(-1+i M_ {l_3,l_3} K _{l_3,l_3}
\right) 
+K _ {l_2,l_2} \left(i+M_ {l_3,l_3} K _{l_3,l_3}\right)) \nonumber \\
&+2M_ {l_2,l_1} K _ {l_2,l_1} \left(i+M_ {l_3,l_3} K _
{l_3,l_3}+M_ {l_3,l_2} \left(K _{l_3,l_2}+i M_ {l_3,l_1}
K _ {l_2,l_1}
K _{l_3,l_3}\right)\right), \nonumber \\
\label{eq_numerator}
\end{eqnarray}
\begin{eqnarray}
\Delta_{D}=&-&2 M_{l_2,l_1} M_ {l_3,l_1}
\left(i K _{l_3,l_2}+M_ {l_3,l_2} K _ {l_3,l_2}^2-M_{l_3,l_2} K
_ {l_2,l_2} K _{l_3,l_3}\right) \nonumber \\
&+&M_ {l_2,l_1}^2 \left(M_ {l_3,l_3} K _ {l_3,l_2}^2-K _ {l_2,l_2} \left(i+M_ {l_3,l_3} K
_{l_3,l_3}\right)\right) \nonumber \\
&+&M_ {l_3,l_1}^2 \left(-i K _
{l_3,l_3}+M_ {l_2,l_2} \left(K _ {l_3,l_2}^2-K _
{l_2,l_2} K _{l_3,l_3}\right)\right) \nonumber \\
&+&M_ {l_1,l_1}
(-1+2 i M_ {l_3,l_2} K _{l_3,l_2}+i M_ {l_3,l_3}
K _{l_3,l_3}+M_ {l_3,l_2}^2 \left(K _ {l_3,l_2}^2-K _ {l_2,l_2}
K _{l_3,l_3}\right) \nonumber \\
&+&M_ {l_2,l_2} \left(-M_{l_3,l_3}
K _ {l_3,l_2}^2+K _ {l_2,l_2} \left(i+M_ {l_3,l_3}
K _{l_3,l_3}\right)\right)), \nonumber \\
\label{eq_denominator}
\end{eqnarray}
and $M_{l,l^{\prime}}$ is the trace $\sum_{n}U_{l,n}^{T}U_{n,l^{\prime}}$ over all the closed transverse harmonic oscillator modes, and are known analytically \cite{pra:89:052716}. The explicit expressions for $M_{l,l^{\prime}}$ are given in the appendix for the cases of identical bosons and fermions considered in this work.

In the dipolar bound state equations (\ref{eq_BoundStateEquation_threeL}),
two sets of quantities are needed to determine the DCIM state. One set is the  $M_{l,l^{\prime}}$, which contain the geometrical information of the waveguide and are known analytically (see appendix). The other set is the elements of the  free-space $\bf{K}^{3D}$ matrix, or equivalently the generalized scattering length $a_{l,l^{\prime}}$ which encapsulate the effect of the interparticle interaction. Eq.~(\ref{eq_BoundStateEquation_threeL}) provides us with the spectrum of the DCIM states once the free-space $\bf{K}^{3D}$ is known. This is one of the main results of this paper.

By setting $K _{l_2,l_3}=K _{l_3,l_2}=K _{l_3,l_3}=0$ in Eq.~(\ref{eq_numerator}) and (\ref{eq_denominator}), one can obtain the bound state equation including two partial wave states. By setting all the other elements of the $\bf{K}^{3D}$ matrix to zero except $K_{l_1,l_1}$, the bound state equation with a single partial wave state is obtained.
\subsubsection{Dipolar bound state equation within Born approximation}
One can obtain the $a_{l,l^{\prime}}$ by solving the free space scattering problem numerically \cite{njp:11:055039}. Alternatively, the Born approximation can be adopted to compute analytically $a_{l,l^\prime}$ away from resonances \cite{njp:11:055039}
\begin{equation}
a_{l,l}=-\frac{2l_d}{(2l-1)(2l+3)},
\end{equation}
and
\begin{equation}
a_{l,l-2}=-\frac{l_d}{(2l-1)\sqrt{(2l+1)(2l-3)}}.
\end{equation}
We note that the Born approximation can not be used to calculate the term $a_{ss}$ \cite{njp:11:055039}.

Applying the Born approximation to calculate the high partial wave elements of the $\bm{K}^{3D}$ matrix, the dipolar bound state equation (\ref{eq_BoundStateEquation_threeL}) for the bosonic dipoles is simplified to
\begin{equation}
a_{ss}=\frac{i}{k}\frac{-1+\eta_1^Bl_d+\eta_2^Bl_d^2+\eta_3^Bl_d^3}{\sigma_0^B+\sigma_1^Bl_d+\sigma_2^Bl_d^2},
\label{eq_BA_boson}
\end{equation}
where
\begin{eqnarray}
\eta_1^B&=&\frac{2 i k}{1155}\left(55 M_{dd}+77 \sqrt{5} M_{ds}+11 \sqrt{5} M_{gd}+15 M_{gg}\right), \nonumber  \\
\eta_2^B&=&\frac{k^2}{3465}\left(77 M_ {ds}^2-7 M_ {gd}^2+2 M_ {ds} \left(11 M_{gd}+6 \sqrt{5} M_{gg}\right) \right.\nonumber \\
&&-\left.12 \sqrt{5} M_{gd} M_{gs}+M_ {dd}\left(7 M_{gg}-22 M_{gs}-77 M_{ss}\right)\right) , \nonumber \\
\eta_3^B&=&-\frac{2 i k^3}{3465}\left(M_ {ds}^2 M_{gg}-2 M_{ds}M_{gd} M_{gs}+M_ {gd}^2 M_{ss}
+M_ {dd} \left(M_ {gs}^2-M_{gg} M_{ss}\right)\right), \nonumber \\
\label{para_boson_numerator}
\end{eqnarray}
and
\begin{eqnarray}
\sigma_0^B&=&-M_{ss}, \nonumber \\
\sigma_1^B&=&-\frac{2 i k}{1155}\left(55 M_ {ds}^2+11 \sqrt{5} M_{ds} M_{gs}+15 M_ {gs}^2-\left(55 M_{dd}+11
\sqrt{5} M_{gd}+15 M_{gg}\right) M_{ss}\right), \nonumber \\
\sigma_2^B&=&-\frac{k^2}{495} \left(M_ {ds}^2 M_{gg}-2 M_{ds}
M_{gd} M_{gs}+M_ {gd}^2 M_{ss}+M_ {dd} \left(M_ {gs}^2-M_{gg} M_{ss}\right)\right).\nonumber \\
\end{eqnarray}
The dipolar bound state equation for the $p$-wave dominated fermionic DCIM state, which exists in the vicinity of free space resonance of $a_{pp}$, is simplified within the Born approximation to
\begin{equation}
a_{pp}=\frac{i}{k}\frac{-1+\eta_1^Fl_d+\eta_2^Fl_d^2+\eta_3^Fl_d^3}{\sigma_0^F+\sigma_1^Fl_d+\sigma_2^Fl_d^2},
\label{eq_BA_fermion}
\end{equation}
where
\begin{eqnarray}
\eta_1^F&=&\frac{2 i k }{45045}\left(1001 M_{ff}+429 \sqrt{21} M_{fp}
+65 \sqrt{77} M_{hf}+385 M_{hh}\right), \nonumber \\
\eta_2^F&=&\frac{k^2}{675675}\left(1287 M_ {fp}^2-405 M_ {hf}^2+10 M_ {fp} \left(13 \sqrt{33} M_{hf}+22 \sqrt{21} M_{hh}\right) \right. \nonumber\\
&&-220\left.\sqrt{21} M_{gd} M_{gs}+M_ {dd} \left(405 M_{hh}-130 \sqrt{33} M_{hp}-1287 M_{pp}\right)\right), \nonumber \\
\eta_3^F&=&-\frac{2	i k^3}{61425}
\left(M_ {fp}^2 M_{hh}-2 M_{fp} M_{hf} M_{hp}+M_ {hf}^2 M_{pp}+M_ {ff} \left(M_ {hp}^2-M_{hh}
M_{pp}\right)\right), \nonumber \\
\label{para_fermion_numerator}
\end{eqnarray}
and
\begin{eqnarray}
\sigma_0^F&=&-M_{pp}, \nonumber \\
\sigma_1^F&=&-\frac{2 i k}{45045} (1001 M_ {fp}^2+65 \sqrt{77} M_{fp} M_{hp}
+385 M_ {hp}^2 \nonumber \\
&&- (1001 M_{ff}+65\sqrt{77} M_{hf}+385 M_{hh}) M_{pp}) , \nonumber \\
\sigma_2^F&=&-\frac{3 k^2}{5005} \left(M_ {fp}^2 M_{hh}-2 M_{fp}
M_{hf} M_{hp}+M_ {hf}^2 M_{pp}+M_ {ff} \left(M_ {hp}^2-M_{hh} M_{pp}\right)\right). \nonumber \\
\label{para_fermion_denominator}
\end{eqnarray}
Compared to the dipolar bound state equation (\ref{eq_BoundStateEquation_threeL}), Eqs. (\ref{eq_BA_boson}) and (\ref{eq_BA_fermion}) explicitly reveal the influence of the DDI in determining the energies of DCIM states. The influence of the waveguide is contained in the two sets of parameters $\eta$ and $\sigma$ in Eqs.~(\ref{eq_BA_boson}) and (\ref{eq_BA_fermion}) which are expressed in terms of $M_{ll^{\prime}}$. These equations allow us to investigate the dependence of DCIM states on $a_{ss}$ (bosonic dipoles) or $a_{pp}$ (fermionic dipoles) for fixed DDI analytically, which will be studied in the following section.
\section{Dipolar confinement induced molecular states}
In the following the dipolar confinement induced bound states are investigated. Both the identical bosonic and fermionic dipoles are considered. In each case, two sets of calculations have been performed. In a first set, the dipole moment $d$ is fixed, $C_{10}$ is varied and accordingly the generalized scattering lengths $a_{l,l^{\prime}}$ change. Such a situation can be realized experimentally by tuning the short-range interaction $V_{\rm sr}({\bm r})$ via Feshbach resonances \cite{rmp:82:1225} while keeping the DDI unchanged. We consider the bound state dominated by its lowest partial wave state, i.e.\ the $s$-wave dominated bound state for identical bosonic dipoles or $p$-wave dominated bound state for identical fermionic dipoles. The variation of the binding energy as a function of $a_{ss}$ or $a_{pp}$ is examined. For the second set we fix $C_{10}$ and allow the dipole moment $d$ to vary. This case can be achieved experimentally, for example in the case of electric dipoles, by tuning external electric fields. The dependence of the bound state energy on the DDI, more specifically the dipole length, is studied.
\subsection{Bosonic DCIM states}
\begin{figure}
	\centering
	\begin{minipage}[b]{0.5\textwidth}
		\includegraphics[width=\textwidth]{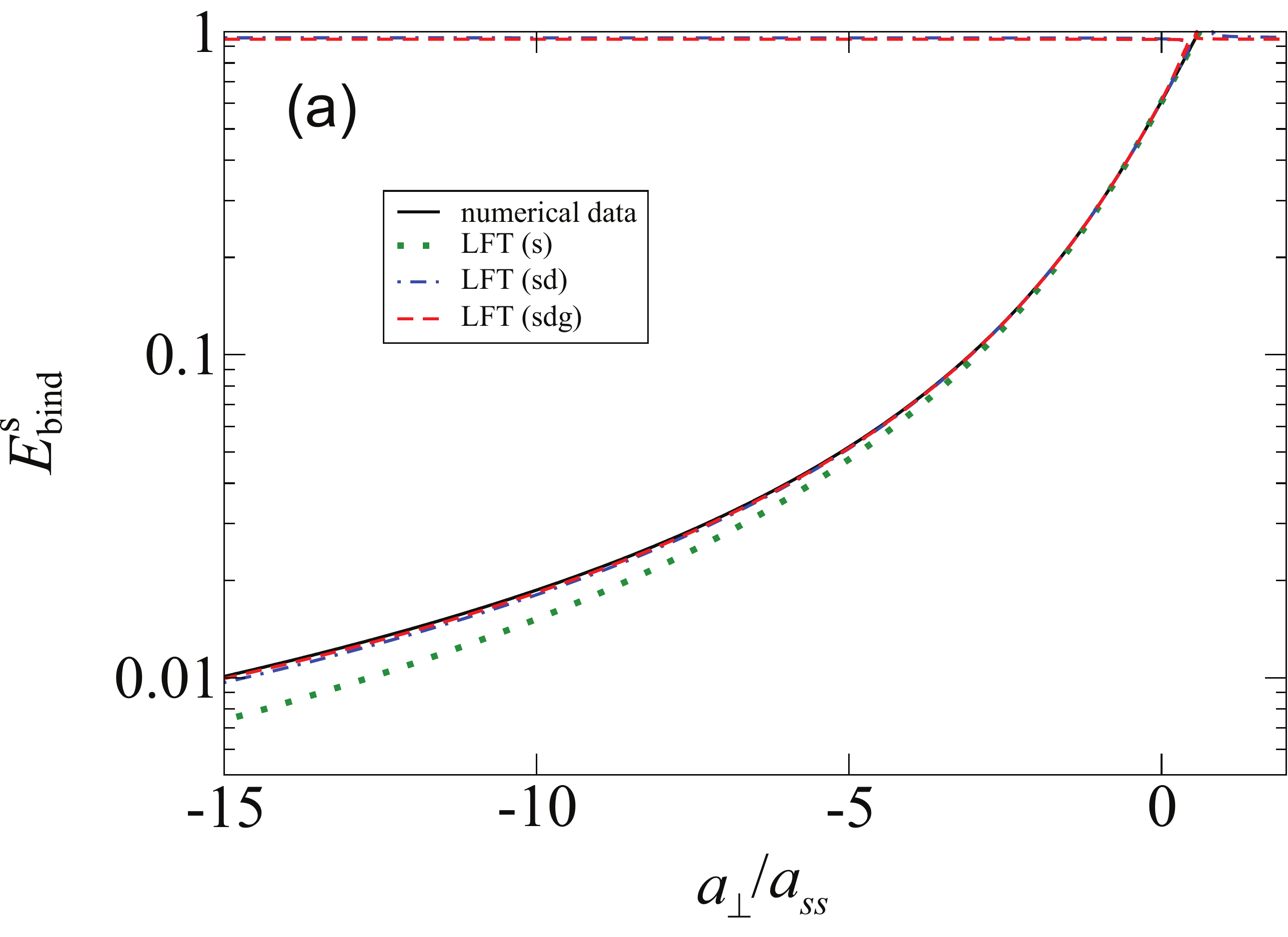} \\
	\end{minipage}
	\begin{minipage}[b]{0.5\textwidth}
		\includegraphics[width=\textwidth]{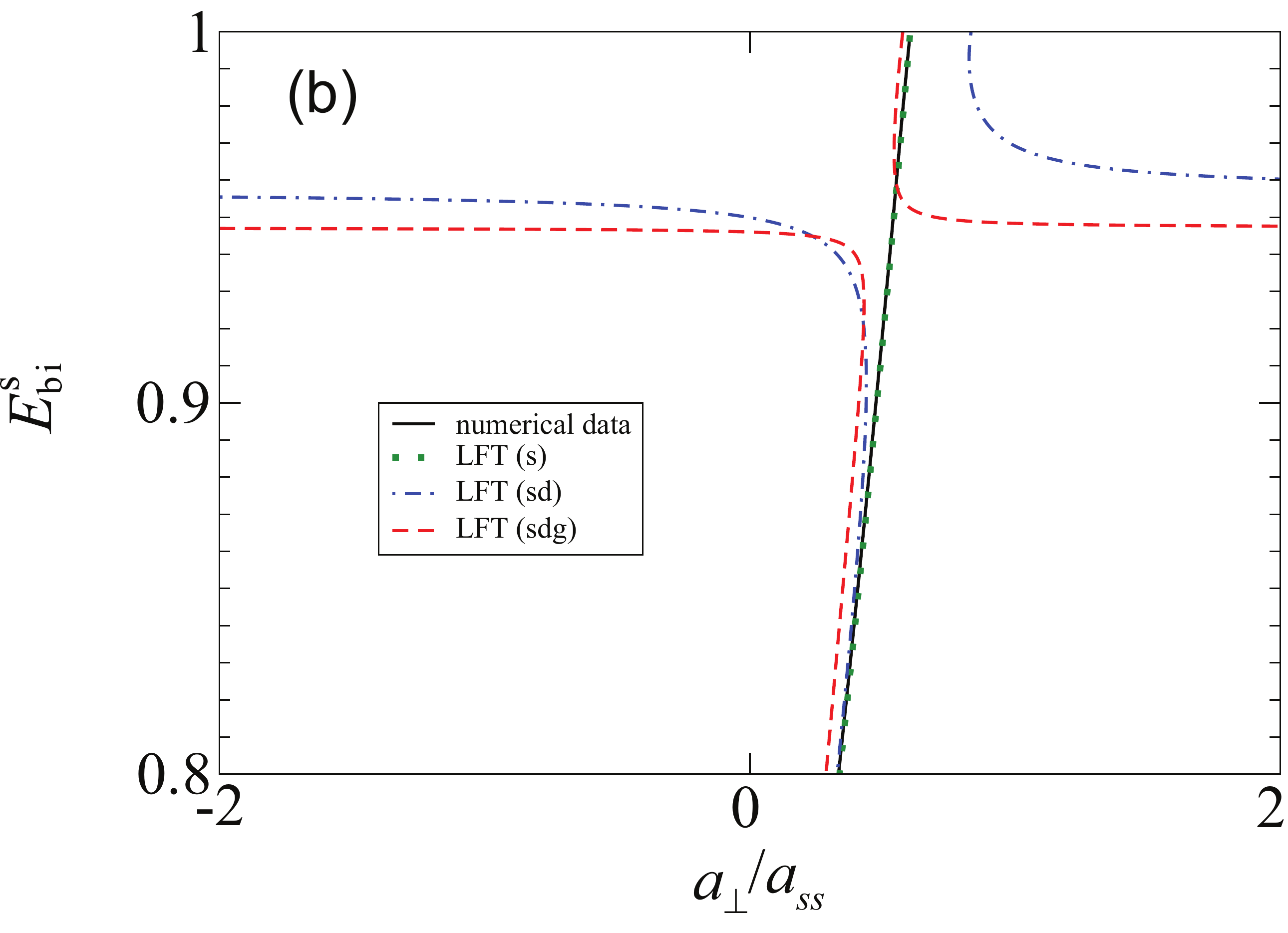}	\\
	\end{minipage}
	\caption{(Upper panel) The scaled binding energy $E_{\rm bi}^s$ of the bosonic DCIM states with varying $a_{\perp}a/_{ss}$. The numerical results (black solid line) are shown together with the LFT calculations including one (green dotted line), two (blue dotted-dashed line) and three (red dashed line) partial wave states. $l_d/a_{\perp}$ is 0.026. (Lower panel) A zoom-in for the energy region where a spurious avoided crossing appears in the LFT calculation including high partial wave states. }
	\label{fig_BosonicBoundStateEnergy}
\end{figure}

For bosonic DCIM states and for a fixed DDI strength $l_d/a_{\perp}=0.026$, the scaled binding energy $E_{\rm bi}^s=E_{\rm bi}/\hbar\omega_{\perp}$ is shown in the upper panel of Fig.~\ref{fig_BosonicBoundStateEnergy} as a function of $a_{\perp}/a_{ss}$.
We note that the binding energy in waveguides is given by $E_{\rm bi}=E_{\rm th}-E_b$.
The numerical data obtained from the close coupling method are shown as a black solid line. $C_{10}$ is varied here in the region where the corresponding LJ potential supports either one or no bound state in free space. We consider the DCIM states in the energy range $0<E_b/\hbar\omega_{\perp}<1$. Once a bound state is determined for a specific $C_{10}$, then the free-space scattering calculation is performed to calculate $a_{ss}$ at the bound state energy.\\
\indent The bound state energy obtained by the LFT approach including one (green dotted line), two (blue dotted-dashed line) and three (red dashed line) partial wave states are also shown in the upper panel of  Fig.~\ref{fig_BosonicBoundStateEnergy}. In this set of LFT calculations, $a_{ss}$ can be treated as a parameter, and all the other generalized scattering lengths needed in the bound state equations can be calculated within the Born approximation. The LFT approach including a single partial wave state reproduces the numerical results well when the binding energy $E_{\rm bi}^s$ is larger than 0.1. Approaching the scattering threshold in the waveguide $E_{\rm bi}^s{\rightarrow}0$, there will be a large portion of the bound state wavefunction spanning over large distances $r/a_{\perp}>1$ where the trapping potential dominates and different partial wave states are strongly coupled together.
In this energy region, the deviation between the bound state energy based on the LFT approach with a single partial wave state and the numerical bound state energy becomes large.
The upper panel of Fig.~\ref{fig_BosonicBoundStateEnergy} shows that including one more partial wave state (blue dotted-dashed line) in the LFT approach can give a more accurate bound state energy below the threshold for the considered regime. Including three partial wave states (red dashed line) improves further the accuracy of the LFT approach in the energy region $E_{\rm bi}^s<0.1$.\\
\indent In the LFT calculation with $s$, $d$ and $s$,$d$,$g$ partial wave states (blue dotted-dashed line and red dashed line in Fig.~\ref{fig_BosonicBoundStateEnergy}), an avoided crossing appears in the binding energy curve in the energy region $E_{\rm bi}^s$ close to one, i.e.\ $E_b$ close to zero, which is not observed in the numerical result and in the LFT approach with $s$ wave state only. The lower panel of Fig.~\ref{fig_BosonicBoundStateEnergy} shows magnification in the vicinity of the avoided crossing. The appearance of the avoided crossing is attributed to the inaccuracy of the local frame transformation for higher partial wave states in the energy region $E_b\rightarrow{0}$. To apply the LFT, an intermediate regime is needed where the kinetic energy dominates the interparticle interaction potential $V_{\rm sr}({\bm r})+V_d({\bm r})$ and the trapping potential $V_{t}({\bm r})$. As stated before we assume an intermediate region exists between $\beta_6/a_{\perp}<r/a_{\perp}<1$. As shown in Fig.~\ref{fig:diabaticPotential}, the channel potentials for higher partial wave states are positive between $\beta_6/a_{\perp}<r/a_{\perp}<1$, while the channel potential energy changes from negative to positive values for the $s$-wave state. If the energy $E$ is close to 0, there is no well-defined intermediate region where the LFT can be applied accurately for higher partial waves. This results in the unphysical avoided crossing in the LFT approach including $d$ and $g$ wave states.
It is worth noting that in the energy region where the avoided crossing appears for higher partial wave states, the bound state energy based on the LFT approach with a single $s$-wave state agrees very well with the numerical results. In addition, these calculation shows that DCIM states also exist in the region $a_{ss}<0$. It demonstrates the impact of the confinement on the dipolar system since in free space case dipolar bound states arise only for $a_{ss}>0$.

\begin{figure}
	\centering
	\begin{minipage}[b]{0.5\textwidth}
		\includegraphics[width=\textwidth]{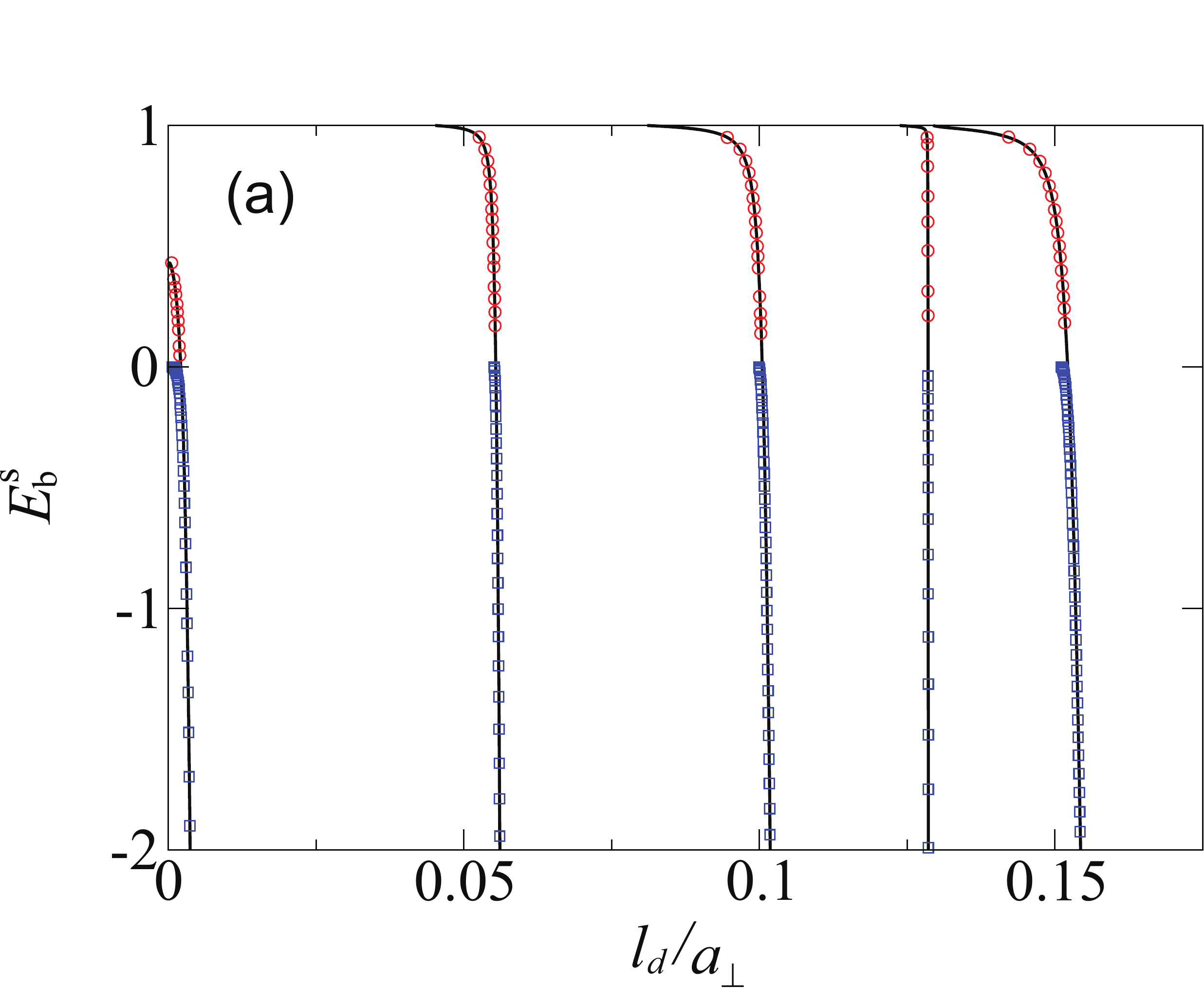} \\
	\end{minipage}
	\begin{minipage}[b]{0.5\textwidth}
		\includegraphics[width=\textwidth]{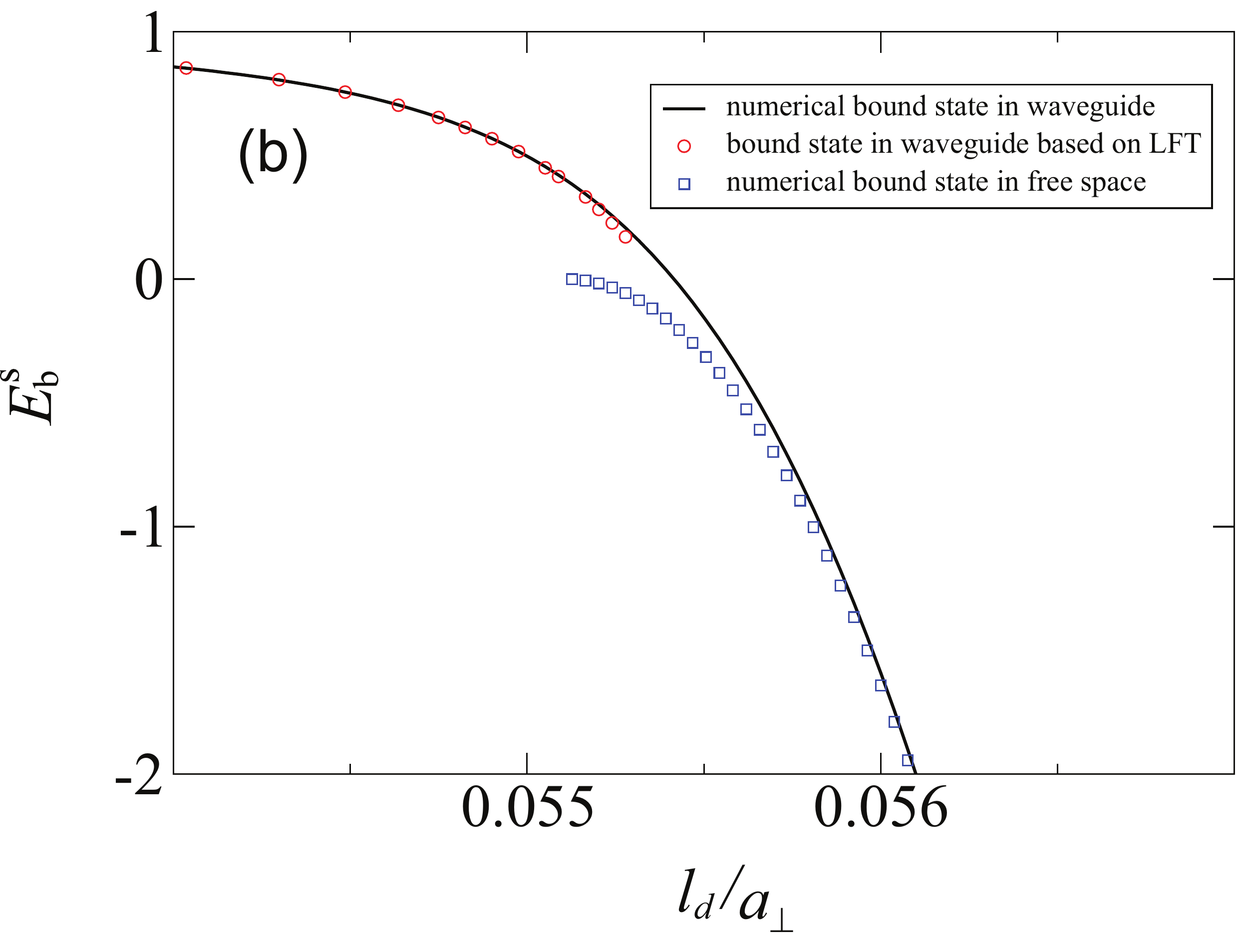}	\\
	\end{minipage}
	\caption{(Upper panel) Black solid line: the numerical scaled energy $E_{b}^s$ of the bosonic DCIM states as a function of scaled dipole length $l_d/a_{\perp}$. Red circles: scaled energies of DCIM states determined within the LFT approach. Blue squares: numerical bound state energy of two identical bosonic dipoles in free space. $C_{10}^s=2.18{\times}10^{-16}$ and the LJ potential supports one bound state which is very close to the threshold in free space.
		(Lower panel) A magnification in the region of the second bound state from the left in the upper panel. }
	\label{fig_BosonicEbVSDipole}
\end{figure}

In the following, we investigate the dependence of the bosonic DCIM states on the DDI strength, characterized by the scaled dipole length $l_d/a_\perp$. The bound state spectrum in the waveguide is calculated numerically via the close-coupling method, and the scaled bound state energy $E_{b}^s=E_b/\hbar\omega_{\perp}$ is shown in Fig.~\ref{fig_BosonicEbVSDipole} (black solid line). The bound state energies based on the LFT approach are shown as red circles. The bound state energies in free space without the transverse trapping potential are also obtained numerically and are depicted as blue squares.
In the numerical calculation, $C_{10}^s$ is fixed to 2.18$\times{10^{-16}}$. The LJ potential supports one bound state which is close to $E=0$, the scattering threshold in free space. Varying the coefficient $C_{10}$, the general features of $E_{b}^s$ as a function of $l_d/a_\perp$ remain the same as those shown in Fig.~\ref{fig_BosonicEbVSDipole}. In the LFT calculation, the dipolar bound state equation~(\ref{eq_BoundStateEquation_threeL}) including three partial waves, which is quite accurate close to the threshold $E_{\rm th}$, is used to determine the bound state in the energy region $E_b/\hbar\omega_\perp\in(1/2,1)$. Away from the threshold, the bound state equation (\ref{eq_BoundStateEquation_threeL}) can be less accurate as shown in Fig.~\ref{fig_BosonicBoundStateEnergy}. Hence, we use the bound state equation including one partial wave state in the energy region $E_b/\hbar\omega_\perp\in(0,1/2)$. The generalized scattering lengths needed in the bound state equations are calculated numerically. As shown in the upper panel of Fig.~\ref{fig_BosonicEbVSDipole}, the bound state energy determined via the LFT approach agrees very well with the numerical results.

As the DDI increases, the interparticle interaction potential becomes deeper \cite{pra:66:052718}. As a result, new bound states emerge as shown in the upper panel of Fig.~\ref{fig_BosonicEbVSDipole}. 
The bound state around $l_d/a_\perp=0.13$ is more sensitive to the variation of the DDI compared to the other bound states. An analysis of the wavefunction reveals that this is a $g$-wave dominant bound state, and all the other bound states are dominated by a $s$-wave component.

As shown in Fig.~\ref{fig_BosonicEbVSDipole}, the bound state energy decreases as the DDI strength increases. When the bound state energy is well below the threshold $E_{\rm th}$, the dipoles are localized at distances $r/a_{\perp}\ll{1}$, and the dipolar bound state wavefunction exponentially vanishes prior to the region where $V_{t}({\bm r})$ becomes important.
Therefore, in this case the bound state in waveguides is essentially like a bound state in free space. This situation is examined in Fig.~\ref{fig_BosonicEbVSDipole} by comparing the scaled bound state energy in the waveguide (black solid line) with the scaled bound state energy in free space (blue squares). In waveguides, the bound state ends at $E=\hbar\omega_\perp$ which is the scattering threshold. In contrast, the scattering threshold in free space is $E=0$, and hence the free-space bound state exists in the energy region $E_{b}^s<0$. We note that the DCIM state considered in this work refers to the bound state in the waveguide and is not necessarily an additional state induced by the confinement. It is clearly shown in Fig.~\ref{fig_BosonicEbVSDipole} that, due to the presence of the confinement, the DCIM state exists in larger energy region and dipole moment region compared to the free-space dipolar bound state. As shown in the upper panel of Fig.~\ref{fig_BosonicEbVSDipole}, the bound state energies with and without trapping potential coincide with each other except for the energy region $E_{b}^s$ close to 0. A magnification in the vicinity of $l_d/a_{\perp}=0.055$ is shown in the lower panel. The bound state which is close to $E=0$ in free space is shifted in the presence of the waveguides. As its energy lowers, this shift becomes smaller. When $E_{b}^s=-1$, i.e. the bound state energy is $-\hbar\omega_{\perp}$, the effect of the waveguide is already negligible, and the bound state energies in the waveguide and in free space are almost the same.
\begin{figure}[!h]\centering
	\resizebox{0.5\textwidth}{!}{
		\includegraphics{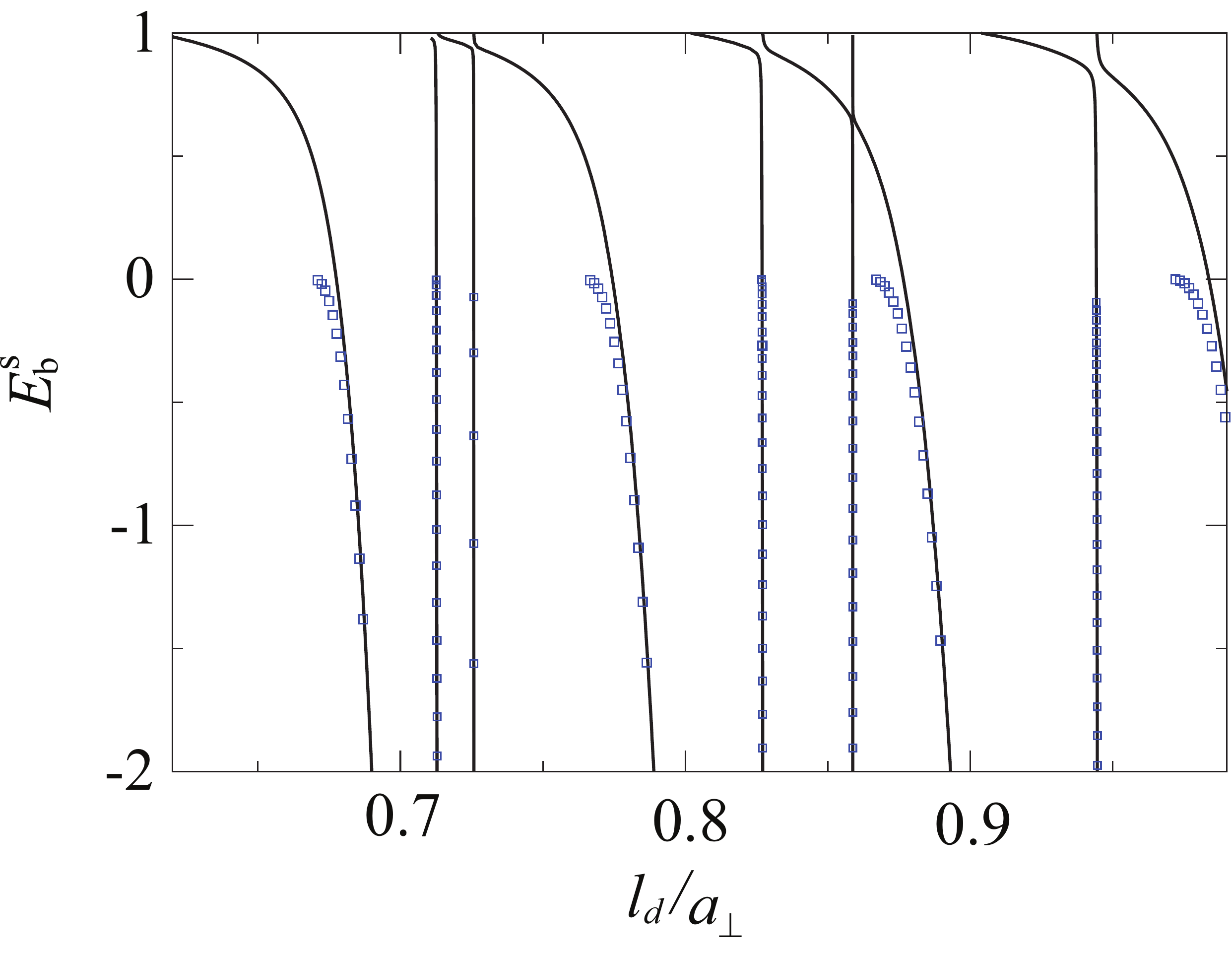}
	} \caption{Black solid line: The numerically obtained scaled energy $E_{b}^s$ of the bosonic DCIM states as a function of the scaled dipole length $l_d/a_{\perp}$ in the strong DDI regime. Blue squares: the scaled bound state energies in free space. $C_{10}^s=2.18{\times}10^{-16}$ and the LJ potential supports one bound state which is very close to the threshold in free space.
}\label{fig:strongDDI}
\end{figure}

In the strong DDI regime, where the dipole length $l_d$ is comparable to the harmonic oscillator length $a_{\perp}$, the dependence of the scaled bound energy $E_{b}^s$ of the DCIM states on $l_d/a_{\perp}$ is determined numerically and is depicted in Fig.~\ref{fig:strongDDI} (black solid line). The LFT approach can not be applied to calculate the DCIM states in the strong DDI regime since the length scale separation does not apply.
The numerical free-space dipolar bound state energies are also shown in Fig.~\ref{fig:strongDDI} (blue squares). 
In the strong DDI regime, different partial wave channels are strongly mixed. As shown in Fig.~\ref{fig:strongDDI}, there are more bound states which are dominated by higher partial wave components and are sensitive to the variation of the DDI strength. Moreover, unlike the situation in the weak DDI regime, where the DCIM states are dominated by a single partial wave component, the DCIM states in the strong DDI regime contain a  significant number of different partial wave components of the same order of magnitude. For example, the first DCIM state from the left in Fig.~\ref{fig:strongDDI} contains both $s$ and $d$ wave components significantly.
The qualitative features of the dependence of the DCIM states on the DDI in the strong DDI regime are similar to those in the weak DDI regime. New DCIM states emerge as the DDI strength increases. In the following discussion on the fermionic DCIM states, we will focus on the weak DDI regime. 

\subsection{Fermionic DCIM states}
\begin{figure}
	\centering
	\begin{minipage}[b]{0.5\textwidth}
		\includegraphics[width=\textwidth]{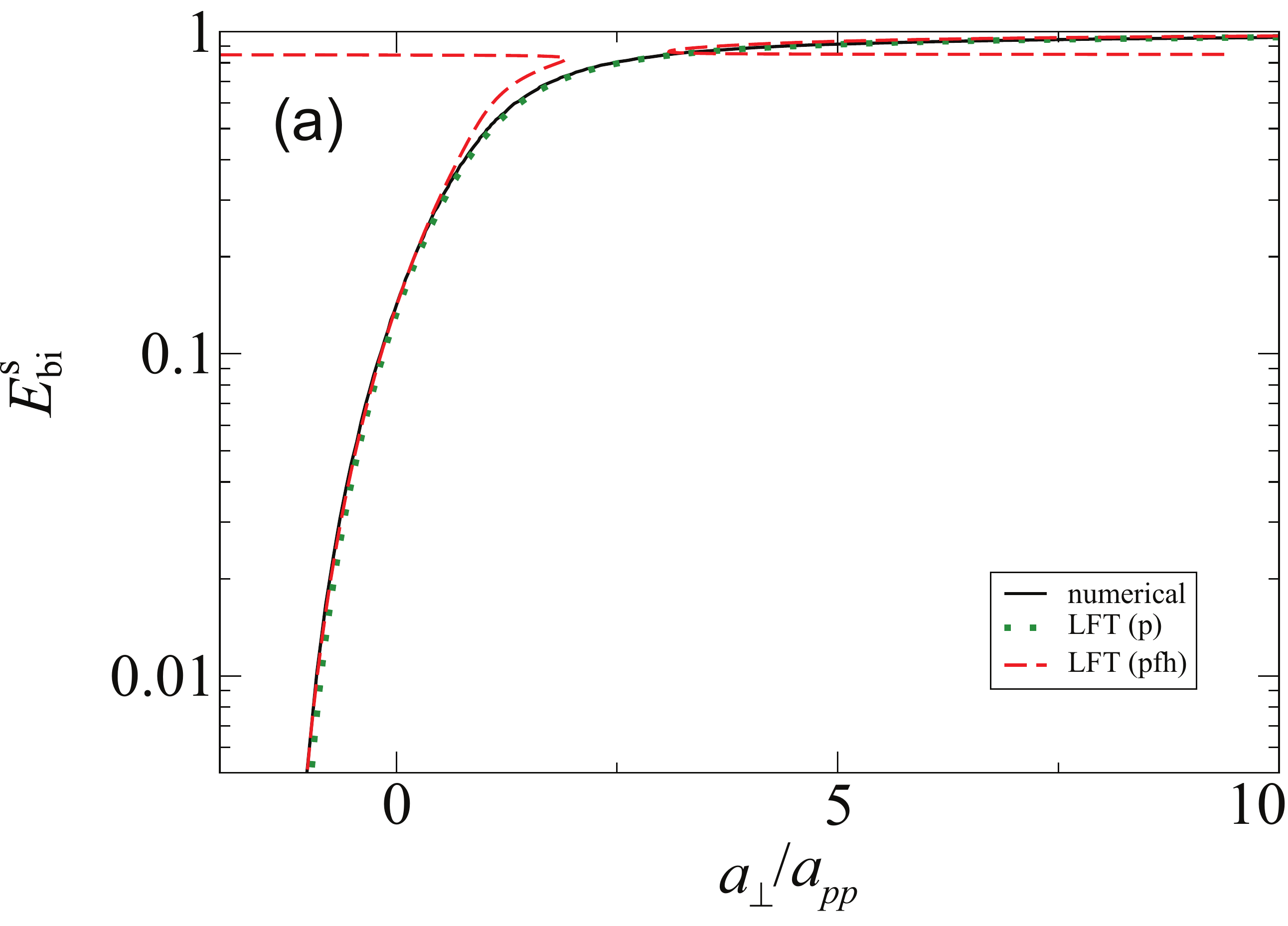} \\
	\end{minipage}
	\begin{minipage}[b]{0.5\textwidth}
		\includegraphics[width=\textwidth]{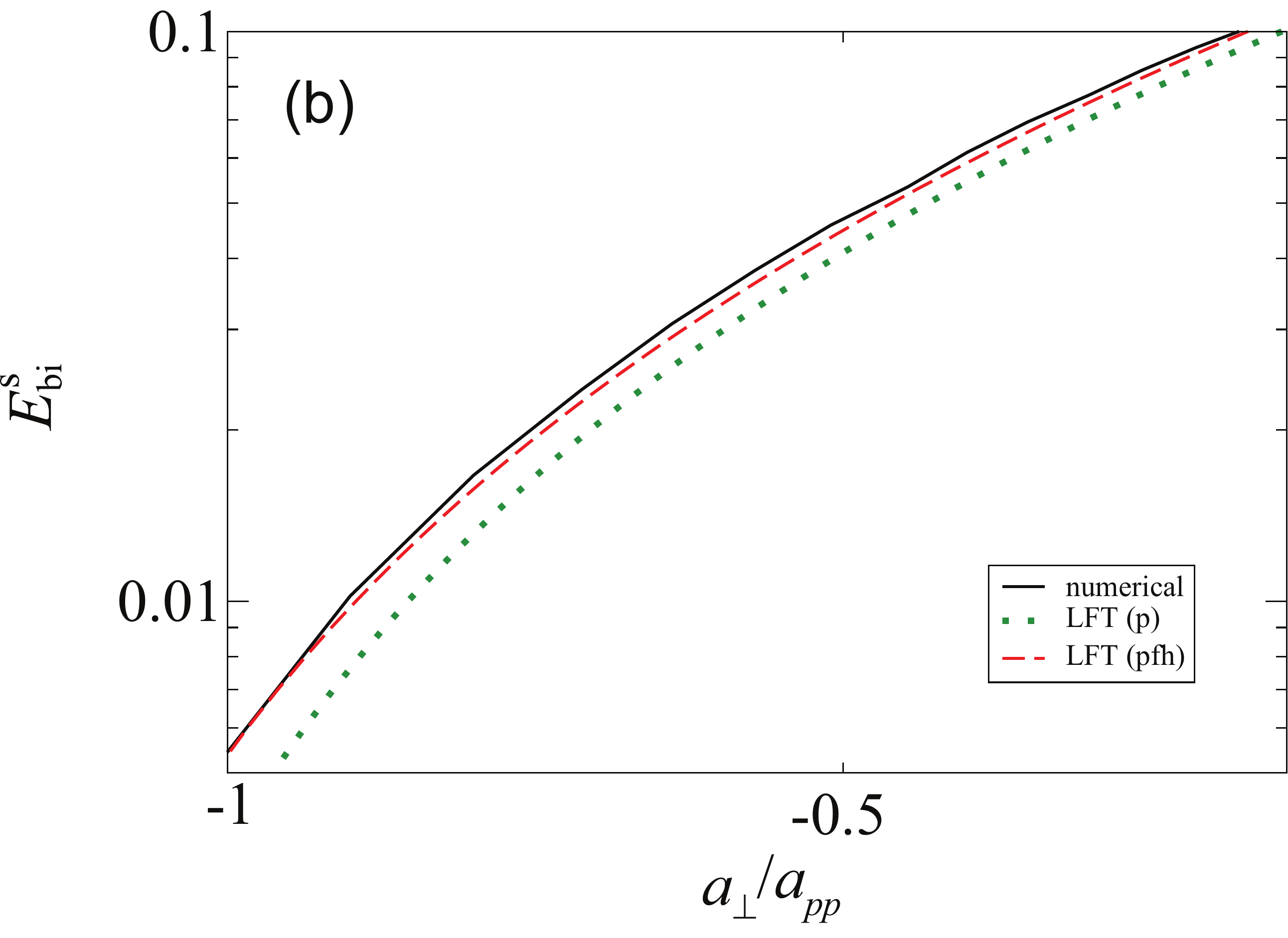}	\\
	\end{minipage}
	\caption{(Upper panel) The scaled binding energy $E_{\rm bi}^s$ of the fermionic DCIM states as a function of $a_{\perp}/a_{pp}$. The numerical results (black solid line) are shown together with the LFT results including one (green dotted line) and three (red dashed line) partial wave states. $l_d/a_{\perp}$ is 0.026. (Lower panel) A magnification in the energy region close to the threshold.}
	\label{fig_FermionicBoundStateEnergy}
\end{figure}

\begin{figure}
	\centering
	\begin{minipage}[b]{0.5\textwidth}
		\includegraphics[width=\textwidth]{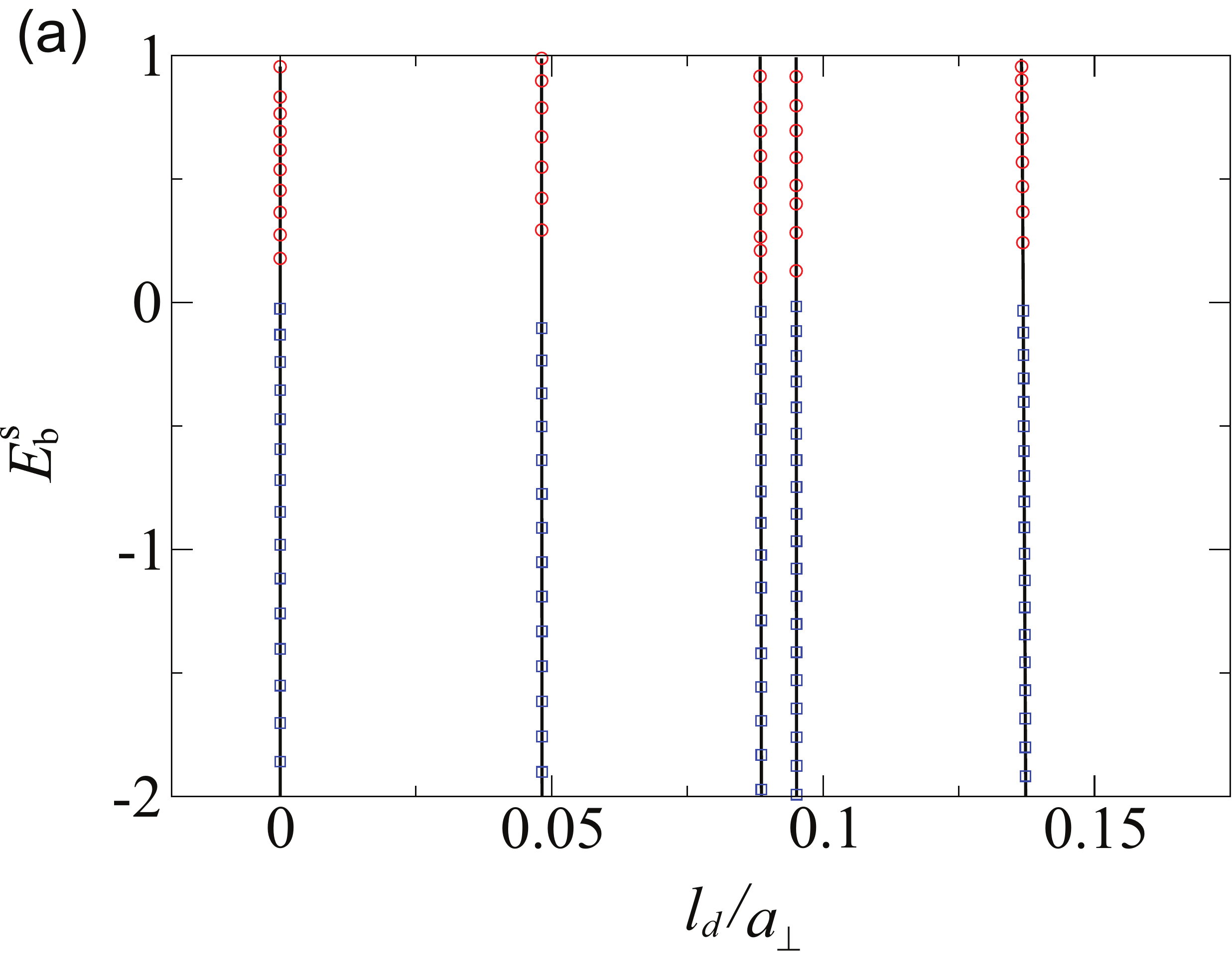} \\
	\end{minipage}
	\begin{minipage}[b]{0.5\textwidth}
		\includegraphics[width=\textwidth]{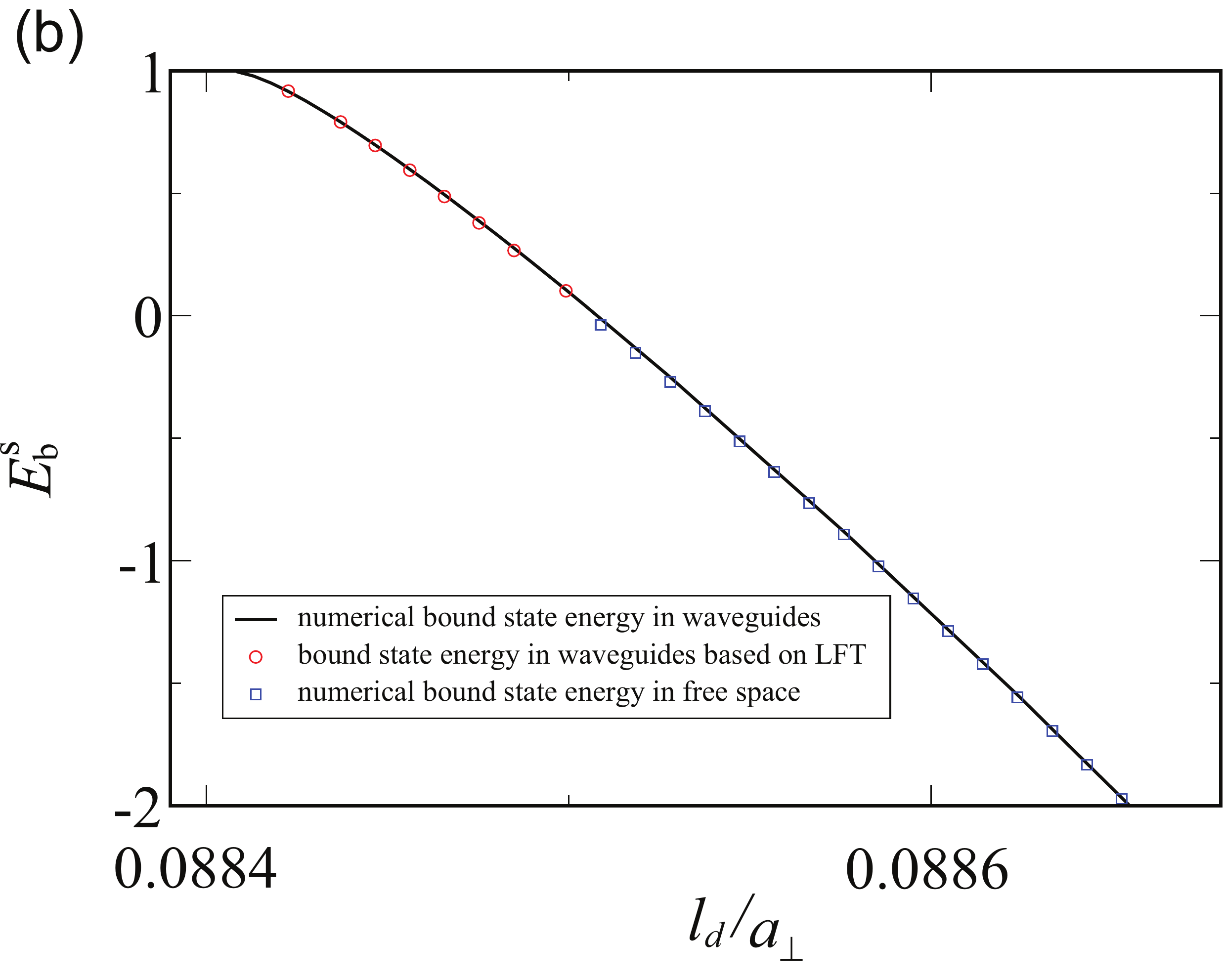}	\\
	\end{minipage}
	\caption{(Upper panel) Black solid line: the numerical scaled energy $E_{b}^s$ of the fermionic DCIM states as a function of the scaled dipole length $l_d/a_{\perp}$. Red circles: scaled energies of the DCIM states obtained from the LFT approach. Blue squares: numerical bound state energy of two identical fermionic dipoles in free space. $C_{10}^s=1.11{\times}10^{-16}$ and the LJ potential supports one $p$-wave bound state which is very close to the threshold in free space.
		(Lower panel) A magnification of the region of the third bound state from the left.}
	\label{fig_FermionicEbVSDipole}
\end{figure}

\indent Let us now focus on the fermionic DCIM states. For fixed DDI strength $l_d/a_{\perp}=0.026$, the variation of the scaled binding energy with $a_{\perp}/a_{pp}$ is shown in Fig.~\ref{fig_FermionicBoundStateEnergy}. The fermionic bound state energy curve has similar features as the bosonic case. As shown in the corresponding lower panel, the LFT approach using a single partial wave state (green dotted line) deviates from the numerical results (black solid line) only close to the threshold. By including more partial wave states (red dashed line), the LFT approach provides a more accurate bound state energy when $E_{\rm bi}^s$ tends to zero, and also produces the spurious avoided crossing when $E_{\rm bi}^s$ tends to 1, see the upper panel of Fig.~\ref{fig_FermionicBoundStateEnergy}.
The dependence of the scaled fermionic bound state energy $E_{\rm bi}^s$ on the DDI strength, namely $l_d/a_{\perp}$, is provided in Fig.~\ref{fig_FermionicEbVSDipole}. Here $C_{10}^s=1.11\times10^{-16}$ and the short range potential $V_{sr}$ supports a $p$-wave bound state very close to $E=0$. Other choices of the value for $C_{10}$ lead to a similar behavior. The solid line in Fig.~\ref{fig_FermionicEbVSDipole} depicts the numerical bound state energies. A series of new bound states emerges as the DDI strength increases. Among the bound states shown in the upper panel of Fig.~\ref{fig_FermionicEbVSDipole}, the one around $l_d/a_\perp=0.095$ is a $l=5$-wave dominant bound state, and the other states are $p$-wave dominated. All these bound states are more sensitive to the variation of the DDI strength compared to the $s$-wave dominant bound state shown in Fig.~\ref{fig_BosonicEbVSDipole}. This can be understood as follows. The potential matrix element $V_d^{l,l}({\bm r})$ vanishes for the $s$-wave channel, and is nonzero for $l>0$ channels. The $l>0$ wave channel potentials are directly affected by the DDI.
Hence the bound states dominated by higher partial wave components are more sensitive to the variation of the DDI.
The bound state energies calculated via the LFT approach are shown in red circles in Fig.~\ref{fig_FermionicEbVSDipole}. The calculation follows the same procedure as described for the bosonic case. It is shown that the LFT approach is also capable to calculate the bound state energies accurately for the fermionic case.

\begin{figure}[!h]\centering
	\resizebox{0.5\textwidth}{!}{
		\includegraphics{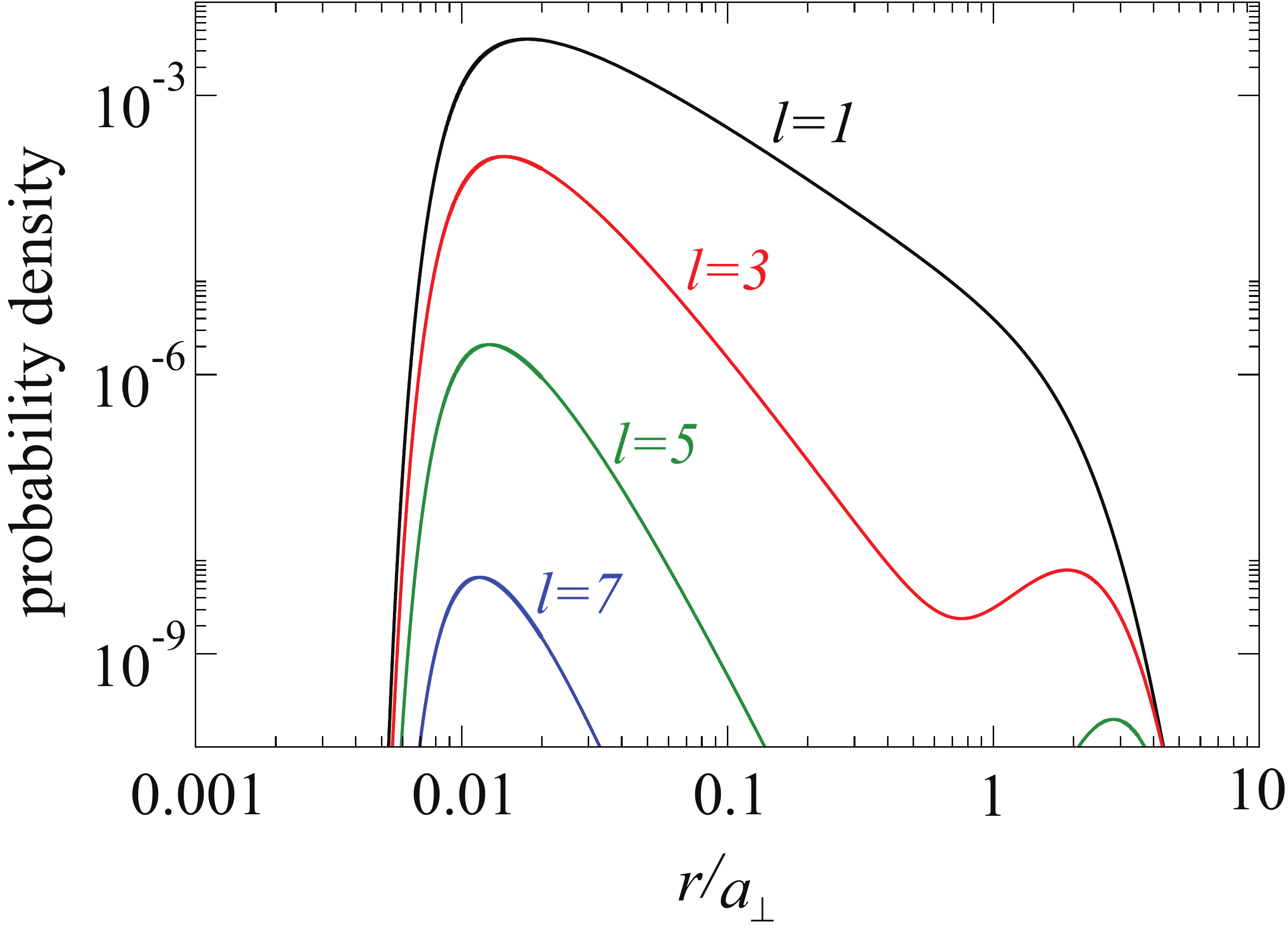}
	} \caption{ The partial wave probability densities of the bound state at $E_{b}^s=0$ for two fermionic dipoles in the waveguide for  $l_d/a_{\perp}=0.026$.
}\label{fig:wv_fermion}
\end{figure}

The numerical free-space bound state energies are shown in blue squares in Fig.~\ref{fig_FermionicEbVSDipole}. Focusing on the bound state in the vicinity of $l_d/a_{\perp}=0.0884$ shown in lower panel, one can observe that, for the fermionic case, the bound state energies in waveguides and in free space coincide with each other even in the energy region $E_{b}^s\rightarrow{0}$. The reason for this is that there are centrifugal potential barriers for all the channels, including the lowest $p$-wave channel. The potential barrier tends to constrain the bound state wavefunction in the short-range region where the trapping potential is negligible. This is clearly shown in Fig.~\ref{fig:wv_fermion} where the partial wave PD of a bound state at $E^s_{b}=0$ in the waveguide is depicted. The bound state barely feels the trapping potential, and is almost a bound state in free space.
\section{Conclusions}
We have investigated the dipolar confinement induced molecular (DCIM) states in harmonic waveguides. Identical bosonic and fermionic dipoles are considered. In the weak DDI regime, in which the dipole length is smaller than the harmonic oscillator length, the local frame transformation (LFT) approach are utilized to connect the bound state in waveguides with the scattering properties in free space analytically. By examining the numerical partial wave probability densities of the DCIM states, we show that length scale separation exists in the weak DDI regime which is crucial for the application of the LFT approach.

The LFT dipolar bound state equation is given. The bound state energies calculated via LFT approach are compared with the numerical ones. Since both DDI and the trapping potential couple different partial wave states, one expects that multiple partial wave states are involved in the LFT approach. Indeed, close to the scattering threshold in waveguides $E=\hbar\omega_{\perp}$, the LFT approach including the lowest partial wave state fails to provide accurate bound state energies, and higher partial wave states are needed. However, when $E$ tends to zero, the bound state energies based on the LFT approach including higher partial wave states deviate from the numerical ones. The reason is that one can not find an intermediate region for higher partial wave channels where the kinetic energy is significantly larger than the interparticle interaction potential and the trapping potential. The local frame transformation in such a case is less accurate, and this results in spurious avoided crossings. Nevertheless, one can still use the LFT approach in this energy region since the single partial wave approximation is valid according to the comparison with the numerical calculations.

The dependence of the DCIM states on the DDI has been investigated, and both the weak and strong DDI regimes have been studied. As the DDI strength increases, a series of DCIM states emerges.
The $s$-wave dominated DCIM states are less sensitive to the variation of the DDI strength as compared to the higher partial wave ($l>0$) dominated DCIM states. This is due to the fact that the matrix elements of the DDI potential vanish for the $s$-wave channel and are nonzero for $l>0$. The $l>0$ channel potentials are affected directly by the DDI.
We also compared the bound states in waveguides and in free space. It is found that for the bosonic case, the bound state in waveguides is almost like a free-space state when the bound state energy is smaller than -$\hbar\omega_{\perp}$. For the fermionic case, the centrifugal potential barrier in the channel potentials localizes the bound state wavefunction in the short-range region where the trapping potential is weak. The fermionic bound states in waveguides and in free space coincide even when energy approaches zero.
\section{Acknowledgments}
G. W. acknowledges a fellowship from the Alexander von Humboldt Foundation.
P. G. acknowledges financial support by the NSF through grant PHY-1306905 and from the Max-Planck Institute for the Physics of Complex Systems in Dresden. G. W. thanks V. Melezhik for fruitful discussions. 
\section{Appendix}
For identical bosons, $s$, $d$ and $g$ wave states are involved, and the explicit expressions for $M_{l,l^{\prime}}$ read as follows
\begin{eqnarray}
M_{ss}=-\frac{i \zeta\left[\frac{1}{2},-\epsilon \right]}{2 \sqrt{\frac{1}{2}+\epsilon }},
\end{eqnarray}
\begin{eqnarray}
M_{ds}=-\sqrt{5} \left(\frac{3 i \zeta\left[-\frac{1}{2},-\epsilon
	\right]}{4 \left(\frac{1}{2}+\epsilon \right)^{3/2}}+\frac{i
	\zeta\left[\frac{1}{2},-\epsilon
	\right]}{4 \sqrt{\frac{1}{2}+\epsilon }}\right),
\end{eqnarray}
\begin{eqnarray}
M_{dd}=5 \left(-\frac{9 i \zeta\left[-\frac{3}{2},-\epsilon
	\right]}{8 \left(\frac{1}{2}+\epsilon \right)^{5/2}}-\frac{3 i
	\zeta\left[-\frac{1}{2},-\epsilon
	\right]}{4 \left(\frac{1}{2}+\epsilon \right)^{3/2}}-\frac{i
	\zeta\left[\frac{1}{2},-\epsilon \right]}{8 \
	\sqrt{\frac{1}{2}+\epsilon }}\right),
\end{eqnarray}
\begin{eqnarray}
M_{gs}=3 \left(-\frac{35 i \zeta\left[-\frac{3}{2},-\epsilon
	\right]}{16 \left(\frac{1}{2}+\epsilon \right)^{5/2}}-\frac{15 i
	\zeta\left[-\frac{1}{2},-\epsilon
	\right]}{8 \left(\frac{1}{2}+\epsilon \right)^{3/2}}-\frac{3 i
	\zeta\left[\frac{1}{2},-\epsilon \right]}{16
	\sqrt{\frac{1}{2}+\epsilon }}\right),
\end{eqnarray}
\begin{eqnarray}
M_{gd}=-3 \sqrt{5} \left(\frac{105 i \zeta\left[-\frac{5}{2},-\epsilon
	\right]}{32 \left(\frac{1}{2}+\epsilon \right)^{7/2}}+\frac{125 i
	\zeta\left[-\frac{3}{2},-\epsilon
	\right]}{32 \left(\frac{1}{2}+\epsilon \right)^{5/2}}+\frac{39 i
	\zeta\left[-\frac{1}{2},-\epsilon \right]}{32 \left(\frac{1}{2}+
	\epsilon \right)^{3/2}}+\frac{3
	i \zeta\left[\frac{1}{2},-\epsilon \right]}{32 \
	\sqrt{\frac{1}{2}+\epsilon }}\right),\nonumber \\
\end{eqnarray}
\begin{eqnarray}
M_{gg}=&9 \left(-\frac{1225 i \zeta\left[-\frac{7}{2},-\epsilon
	\right]}{128 \left(\frac{1}{2}+\epsilon \right)^{9/2}}\right.-\frac{525 i
	\zeta\left[-\frac{5}{2},-\epsilon
	\right]}{32 \left(\frac{1}{2}+\epsilon \right)^{7/2}}-\frac{555 i
	\zeta\left[-\frac{3}{2},-\epsilon \right]}{64 \left(\frac{1}{2}+
	\epsilon
	\right)^{5/2}} \nonumber \\
&-\frac{45 i \zeta\left[-\frac{1}{2},-\epsilon
	\right]}{32 \left(\frac{1}{2}+\epsilon \right)^{3/2}}
-\left.\frac{9 i
	\zeta\left[\frac{1}{2},-\epsilon
	\right]}{128 \sqrt{\frac{1}{2}+\epsilon }}\right),
\end{eqnarray}
where $\epsilon=\frac{E-\hbar*\omega_{\perp}}{2\hbar\omega_{\perp}}$, and
$E$ is the total energy. $\zeta(a,s)$ is the Hurwitz zeta function.\\
For identical fermions, $p$, $f$ and $h$ waves are involved in the LFT calculation. The explicit expressions for $M_{l,l^{\prime}}$ read then
\begin{eqnarray}
M_{pp}=\frac{3 i \zeta\left[-\frac{1}{2},-\epsilon \right]}{2 \left(\frac{1}{2}+\epsilon \right)^{3/2}},
\end{eqnarray}
\begin{eqnarray}
M_{fp}=-\sqrt{21} \left(-\frac{5 i \zeta\left[-\frac{3}{2},-\epsilon \right]}{4 \left(\frac{1}{2}+\epsilon \right)^{5/2}}-\frac{3 i \zeta\left[-\frac{1}{2},-\epsilon
	\right]}{4 \left(\frac{1}{2}+\epsilon \right)^{3/2}}\right),
\end{eqnarray}
\begin{eqnarray}
M_{ff}=7 \left(\frac{25 i \zeta\left[-\frac{5}{2},-\epsilon \right]}{8 \left(\frac{1}{2}+\epsilon \right)^{7/2}}+\frac{15 i \zeta\left[-\frac{3}{2},-\epsilon
	\right]}{4 \left(\frac{1}{2}+\epsilon \right)^{5/2}}+\frac{9 i \zeta\left[-\frac{1}{2},-\epsilon \right]}{8
	\left(\frac{1}{2}+\epsilon \right)^{3/2}}\right),
\end{eqnarray}
\begin{eqnarray}
M_{hp}=\sqrt{33} \left(\frac{63 i \zeta\left[-\frac{5}{2},-\epsilon \right]}{16 \left(\frac{1}{2}+\epsilon \right)^{7/2}}+\frac{35 i \zeta\left[-\frac{3}{2},-\epsilon
	\right]}{8 \left(\frac{1}{2}+\epsilon \right)^{5/2}}+\frac{15 i \zeta\left[-\frac{1}{2},-\epsilon \right]}{16 \left(\frac{1}{2}+
	\epsilon \right)^{3/2}}\right),
\end{eqnarray}
\begin{eqnarray}
M_{hf}=-\sqrt{77} \left(-\frac{315 i \zeta\left[-\frac{7}{2},-\epsilon \right]}{32 \left(\frac{1}{2}+\epsilon \right)^{9/2}}-\frac{539 i \zeta\left[-\frac{5}{2},-\epsilon
	\right]}{32 \left(\frac{1}{2}+\epsilon \right)^{7/2}}-\frac{285 i \zeta\left[-\frac{3}{2},-\epsilon \right]}{32 \left(\frac{1}{2}+\epsilon
	\right)^{5/2}}-\frac{45 i \zeta\left[-\frac{1}{2},-\epsilon \right]}{32 \left(\frac{1}{2}+\epsilon \right)^{3/2}}\right),\nonumber \\
\end{eqnarray}
\begin{eqnarray}
M_{hh}=&11 \left(\frac{3969 i \zeta\left[-\frac{9}{2},-\epsilon \right]}{128 \left(\frac{1}{2}+\epsilon \right)^{11/2}}+\frac{2205 i \zeta\left[-\frac{7}{2},-\epsilon
	\right]}{32 \left(\frac{1}{2}+\epsilon \right)^{9/2}}+\frac{3395 i \zeta\left[-\frac{5}{2},-\epsilon \right]}{64 \left(\frac{1}{2}+\epsilon
	\right)^{7/2}} \right.\nonumber \\
&+\left.\frac{525 i \zeta\left[-\frac{3}{2},-\epsilon \right]}{32 \left(\frac{1}{2}+\epsilon \right)^{5/2}}+\frac{225 i \zeta\left[-\frac{1}{2},-\epsilon
	\right]}{128 \left(\frac{1}{2}+\epsilon \right)^{3/2}}\right).
\end{eqnarray}



\end{document}